\documentclass[12pt]{article}%
\usepackage{amsfonts}
\usepackage{amssymb}
\usepackage{amsmath}
\usepackage{geometry}
\usepackage{graphicx}%
\providecommand{\U}[1]{\protect\rule{.1in}{.1in}}

\usepackage{subcaption}
\usepackage{multirow}
\usepackage{pdflscape}
\newcommand{\ind}{\perp\!\!\!\perp}

\usepackage{pifont}

\usepackage{pgf,tikz,wrapfig}
\usetikzlibrary{arrows,shapes.arrows,shapes.geometric,shapes.multipart,decorations.pathmorphing,positioning}
\tikzset{
	>=stealth',
	true/.style={
		rectangle,
		draw=black, very thick,
		text width=6.5em,
		minimum height=2em,
		text centered,
		fill=gray, opacity = 0.5},
	punkt/.style={
		rectangle,
		rounded corners,
		draw=black, very thick,
		text width=6.5em,
		minimum height=2em,
		text centered},
	est/.style={
		circle,
		draw=black, very thick,
		text centered},
	shade/.style={
		circle,
		draw=black, very thick, fill=gray!50,
		text centered},
	weight/.style={
		circle,
		draw=black, very thick,
		text width=6.5em,
		minimum height=2em,
		text centered},
	pil/.style={
		->,
		thick,
		shorten <=2pt,
		shorten >=2pt,},
	double/.style={
		<->,
		very thick,
		shorten <=2pt,
		shorten >=2pt,},
	dash/.style={
		dashed,
		very thick,
		shorten <=2pt,
		shorten >=2pt,},
	dashdouble/.style={
		<->,
		dashed,
		thick,
		shorten <=2pt,
		shorten >=2pt}}
\begin{document}

\title{ }

\begin{center}
O\textit{riginal Article}\bigskip

{\LARGE An Introduction to Proximal Causal Learning}

\bigskip

{\Large Eric J Tchetgen Tchetgen }

{\Large Andrew Ying }

{\Large Yifan Cui}

{\large Department of Statistics, The Wharton School, University of
Pennsylvania }

\medskip

{\Large Xu Shi }

{\large Department of Biostatistics, University of Michigan }

\medskip

{\Large Wang Miao}

{\large Peking University }

\textbf{Abstract}
\end{center}

\noindent A standard assumption for causal inference from observational data
is that one has measured a sufficiently rich set of covariates to ensure that
within covariate strata, subjects are exchangeable across observed treatment
values. Skepticism about the exchangeability assumption in observational
studies is often warranted because it hinges on investigators' ability to
accurately measure covariates capturing all potential sources of confounding.
Realistically, confounding mechanisms can rarely if ever, be learned with
certainty from measured covariates. One can therefore only ever hope that
covariate measurements are at best proxies of true underlying confounding
mechanisms operating in an observational study, thus invalidating causal
claims made on basis of standard exchangeability conditions. Causal learning
from proxies is a challenging inverse problem which has to date remained
unresolved. In this paper, we introduce a formal potential outcome framework
for \textit{proximal causal learning}, which while explicitly acknowledging
covariate measurements as imperfect proxies of confounding mechanisms, offers
an opportunity to learn about causal effects in settings where exchangeability
on the basis of measured covariates fails. Sufficient conditions for
nonparametric identification are given, leading to the \textit{proximal
g-formula} and corresponding \textit{proximal g-computation}
\textit{algorithm} for estimation. These may be viewed as generalizations of
Robins' foundational g-formula and g-computation algorithm, which account
explicitly\ for bias due to unmeasured confounding. Both point treatment and
time-varying treatment settings are considered, and an application of proximal
g-computation of causal effects is given for illustration.

\noindent


\bigskip\pagebreak

\section{\noindent Introduction}

A key assumption routinely made for causal inference from observational data
is that one has measured a sufficiently rich set of covariates, to ensure that
within covariate strata, subjects are exchangeable across observed treatment
values$^{1,2}$. This fundamental assumption is inherently untestable
empirically, without introducing a different untestable assumption, and
therefore must be taken on faith even\ with substantial subject matter
knowledge at hand. For this reason, the assumption of exchangeability in
observational studies is often the subject of much skepticism, mainly because
it hinges on an assumed ability of the investigator to accurately measure
covariates relevant to the various confounding mechanisms potentially present
in a given observational study. Realistically, confounding mechanisms can
rarely if ever, be learned with certainty from measured covariates. Therefore,
the most one can hope for in practice, is that covariate measurements are at
best proxies of the true underlying confounding mechanism operating in a given
observational study. Such acknowledgement invalidates any causal claim made on
the basis of exchangeability. In this paper, we introduce a general framework
for \textit{proximal causal learning}, which while explicitly acknowledging
covariate measurements as imperfect proxies of confounding mechanisms, enables
one to potentially learn about causal effects in settings where
exchangeability does not hold on the basis of measured covariates.

As all formal methods for causal inference, the proposed proximal approach
relies on assumptions that are not testable empirically without a different
assumption; nevertheless, as we argue next, we view the required identifying
assumptions as easily interpretable and potentially easier to reason about on
subject matter grounds than exchangeability. Mainly, proximal causal learning
requires that the analyst can correctly classify proxies into three bucket
types: a. variables which are common causes of the treatment and outcome
variables; b. treatment-inducing confounding proxies\ versus c.
outcome-inducing confounding proxies. A proxy of type b is a potential cause
of the treatment which is related with the outcome only through an unmeasured
common cause for which the variable is a proxy; while a proxy of type c is a
potential cause of the outcome which is related with the treatment only
through an unmeasured common cause for which the variable\ is a proxy. Proxies
that are neither causes of treatment or outcome variable may belong to either
bucket type b or c. \ Examples of proxies of type b and c abound in
observational studies. For instance, in an observational study evaluating the
effects of a treatment on disease progression, one is typically concerned that
patients either self-select or are selected by their physician to take the
treatment based on prognostic factors for the outcome; therefore there may be
two distinct processes contributing to a subject's propensity to be treated.
In an effort to account for these sources of confounding, a diligent
investigator would endeavor to record lab measurements and other clinically
relevant covariate data available to the physician and the patient when
considering treatment options. For instance, it is customary in evaluating the
effectiveness of HIV\ anti-retroviral therapy, to adjust for CD4 count
measurement as a potential source of confounding by indication$^{3}$; this is
because whenever available to the prescribing physician, CD4 count measurement
is invariably used to decide (at least prior to the advent of universal test
and treat) whether a patient should be given ART, i.e. the probability of
treatment initiation generally decreases with a patient's increasing CD4
count. However, as an error-prone snapshot measurement of the evolving state
of the patient's underlying immune system, CD4 count measurement is unlikely
to be a direct cause of disease progression but rather a proxy of the actual
state of patient's immune system, the actual cause of disease progression. It
is therefore more accurate to conceive of baseline CD4 count measurement as an
imperfect proxy of immune system status.\ In addition, two patients with
similar CD4 count measurements seen by different physicians may differ in
terms of treatment decisions depending on the patient's own health seeking
behavior, as well as differences in physician's clinical training and
experience, and overall prescription preferences; which are all factors
potentially predictive of patient disease progression regardless of treatment.
Because such factors are notoriously difficult to measure accurately, they are
likely to induce residual confounding, even after adjusting for baseline CD4 count.

A proxy of type c might include baseline covariate measurements assessing a
patient's mental and physical commorbidities including those measured using a
validated questionnaire. For instance, it is well known that in addition to
the immune system, HIV affects the nervous system and the brain producing
neurological sequelae, often resulting in forgetfulness and cognitive
problems$^{4,5}$. These problems can compromise medication adherence,
interfere with instrumental activities of daily living such as driving and
managing finances, increase dependency, and decrease quality of life. Several
cognitive functioning screening tools exist to objectively measure cognition,
with the gold standard being the Mini-Mental State Examination (MMSE)$^{6.7}$,
which is a 30-point questionnaire that is used extensively in clinical and
research settings. A score of 24 or less is generally used as a cut-off to
indicate possible mild cognitive impairment or early stage dementia$^{6}$
although the cut-off can vary according to the education level of the
individual$^{8}$.\ Although widely used as a measure of cognition, MMSE\ is at
best an imperfect proxy of a patient's baseline state of cognitive impairment,
which may in turn influence both a patient's willingness to initiate and
adhere to ART, and the patient's disease progression at follow-up. Thus, in
evaluating the causal effect of ART on disease progression such as say
cognitive decline, baseline MMSE measurements can be seen as a proxy of type c
for the underlying confounding mechanism corresponding to the patient's
underlying state of cognitive impairment at baseline. These are but
two\ motivating examples of confounding proxies in analysis of the causal
effects of ART on HIV\ infection related disease progression from an
observational study. Aside for these proxies, there may also be factors that
can accurately be described as true common causes of treatment and outcome
processes; these variables which we have referred to as of type a may in fact
include age, gender and years of education depending on the context. Thus,
rather than as current practice dictates, assuming that adjusting for baseline
covariates, exchangeability can be attained, our proposed proximal framework
requires the investigator correctly classifies covariates that belong in
bucket types a, b and c without necessarily the need for exchangeability to
hold conditional on such proxies.

In order to ground ideas, we briefly describe the proposed proximal approach
in the context of a point exposure $A,$ outcome $Y$, and unmeasured confounder
$U;$ then suppose that one can correctly select a treatment-inducing proxy $Z$
and an outcome-inducing proxy $W$ such that the simple structural linear model
given below holds:%
\begin{align}
E(Y|A,Z,X,U) &  =\beta_{0}+\beta_{a}A+\beta_{u}U+\beta_{x}^{\prime
}X\label{Linear Structural}\\
E(W|A,Z,X,U) &  =\eta_{0}+\eta_{u}U+\eta_{x}^{\prime}X.\nonumber
\end{align}
where $X$ are all other observed covariates, and validity of proxies is
encoded by the fact that the right handside of the first equation does not
depend on $Z$, the right handside of the second equation does not depend on
$A$ and $Z;$ and $W$ is $U$ relevant in the sense that $\eta_{u}\neq0.$ The
causal parameter of interest is $\beta_{a}=E(Y_{a+1}-Y_{a}|U,X)\ $
corresponding to the average outcome difference if one were to intervene to
increase the treatment by one unit upon conditioning on covariates $\left(
U,X\right)  $ a sufficient confounding adjustment set; that is exchangeability
holds conditional on $\left(  U,X\right)  .$ It is then straightforward to
show that
\begin{align*}
E(Y|A,Z,X) &  =\beta_{0}+\beta_{a}A+\beta_{u}E\left(  U|A,Z,X\right)
+\beta_{x}^{\prime}X\\
E(W|A,Z,X) &  =\eta_{0}+\eta_{u}E\left(  U|A,Z,X\right)  +\eta_{x}^{\prime}X,
\end{align*}
so that
\begin{equation}
E(Y|A,Z,X)=\beta_{0}^{\ast}+\beta_{a}A+\beta_{u}^{\ast}E(W|A,Z,X)+\beta
_{x}^{\ast\prime}X\label{proximal deconfounder1}%
\end{equation}
where
\begin{align*}
\beta_{0}^{\ast} &  =\beta_{0}-\frac{\beta_{u}\eta_{0}}{\eta_{u}}\\
\beta_{u}^{\ast} &  =\frac{\beta_{u}}{\eta_{u}}\\
\beta_{x}^{\ast} &  =\beta_{x}-\frac{\beta_{u}\eta_{x}}{\eta_{u}}%
\end{align*}
Let $\widehat{W}$ denote an (asymptotically) unbiased estimator of
$E(W|A,Z,X)$, then equation $\left(  \ref{proximal deconfounder1}\right)  $
suggests that provided that $E\left(  U|A,Z,X\right)  $ depends on $Z,$ the
least squares linear regression of $Y$ on $(A,X,\widehat{W})$ recovers a slope
coefficient for $A,$ $\widehat{\beta}_{a}$ that is consistent for the causal
parameter $\beta_{a}.$ In contrast, either removing $\widehat{W}$ from the
regression model, or replacing it with either $W$ or $(W,Z)$ will generally
yield a biased estimate of $\beta_{a}$ given that exchangeability does not
hold either conditional on $X,$ on ($X,W),$ or on $(X,W,Z).$ As further
adjusting for $\widehat{W}$ debiases the least squares estimator of $\beta
_{a}$ conditional on $X,$ we shall refer to $\widehat{W}$ as a
\textit{proximal control variable}. The system of linear structural equations
considered above is overly restrictive, assuming linearity and no
interactions; as we demonstrate in this paper, these assumptions are not
strictly necessary and can be relaxed considerably so that nonparametric
identification remains possible under certain conditions.

Notation and formal definitions used throughout the paper are given in the
next section, nonparametric identification conditions for proximal causal
learning are presented in the following section, where it is shown that causal
effects can sometimes be identified by the \textit{proximal g-formula,} a
generalization of Robins' foundational g-formula which accounts for
confounding bias due to unmeasured factors. For estimation, a \textit{proximal
g-computation algorithm }is then introduced. As we show, equation $\left(
\ref{proximal deconfounder1}\right)  $ can be recovered as a special case of
proximal g-computation algorithm. Both point treatment and time-varying
treatment settings are considered, and applications of proximal causal
learning are given to illustrate the methodology. The paper concludes with
brief final remarks.

\section{\noindent{\protect\Large Notation and definitions}}

\noindent Suppose one has observed i.i.d samples on $(A,L,Y)$ where $A$
denotes a treatment of interest, $Y$ is an outcome of interest and $L$ is a
set of measured covariates. Let $Y_{a}$ denote the potential outcome had,
possibly contrary to fact a person received treatment $A=a.$ Throughout, we
make the standard consistency assumption linking observed and potential
outcomes%
\begin{equation}
Y=Y_{A}.\label{consistency}%
\end{equation}
We aim to identify a population average causal effect, corresponding to a
contrast of counterfactual averages $\beta\left(  a\right)  =E(Y_{a})$ for
different values of $a.$ For instance, in case of binary treatment, one might
be interested in the average treatment effect measured on the additive scale
$\beta\left(  1\right)  -\beta\left(  0\right)  =E(Y_{1})-E(Y_{0});$ in case
of binary outcome, one might also be interested in the average treatment
effect on the multiplicative scale $\beta\left(  1\right)  /\beta\left(
0\right)  =\Pr(Y_{1}=1)/\Pr(Y_{0}=1).$ In all cases, whether $A$ is binary,
polytomous or continuous, learning about causal effects on any given scale
involves learning about the potential outcome mean $\beta\left(  a\right)  ,$
which we aim to identify from the observed sample. A common identification
strategy in observational studies is that of exchangeability$^{1,2,9}$ or no
unmeasured confounding (NUC) condition on the basis of measured covariates:%
\begin{equation}
Y_{a}\amalg A|L\label{exchangeability}%
\end{equation}
where $\amalg$ denotes independence, together with positivity condition that
\begin{equation}
f(A=a|L)>0\label{positivity}%
\end{equation}
where $f\left(  d|g\right)  $ denotes the conditional density or probability
mass function of $d$ given $g.$ Assumption $\left(  \ref{exchangeability}%
\right)  $ is sometimes interpreted as stating that $L$ includes all common
causes of $A$ and $Y;$ an assumption represented in causal directed acyclic
graph (DAG) in Figure 1.a. in which $L$ is of Type a.%

\begin{figure}[h]
	\begin{minipage}[b]{0.3\linewidth}
		\centering
		\includegraphics[scale=1]{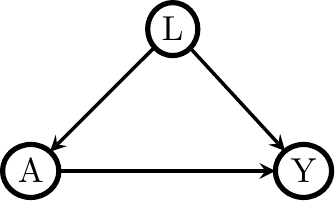}
		\subcaption{Type (a) proxy.}
		\label{fig:dag1a}
	\end{minipage}
	\begin{minipage}[b]{0.3\linewidth}
		\centering
		\includegraphics[scale=1]{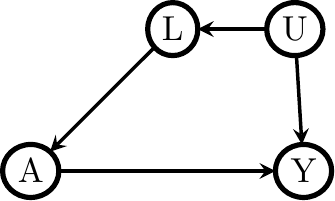}
		\subcaption{Type (b) proxy.}
		\label{fig:dag1b}
	\end{minipage}
	\begin{minipage}[b]{0.3\linewidth}
		\centering
		\includegraphics[scale=1]{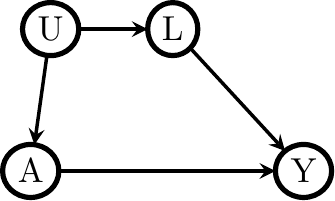}
		\subcaption{Type (c) proxy.}
		\label{fig:dag1c}
	\end{minipage}
	\begin{minipage}[b]{0.45\linewidth}
		\centering
		\includegraphics[scale=1]{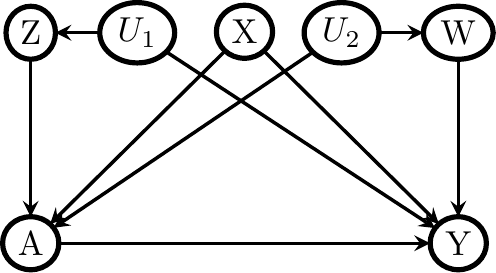}
		\subcaption{Coexistence of type (a)(b)(c) proxies when NUC holds.}
		\label{fig:dag1d}
	\end{minipage}\hfill
	\begin{minipage}[b]{0.45\linewidth}
		\centering
		\includegraphics[scale=1]{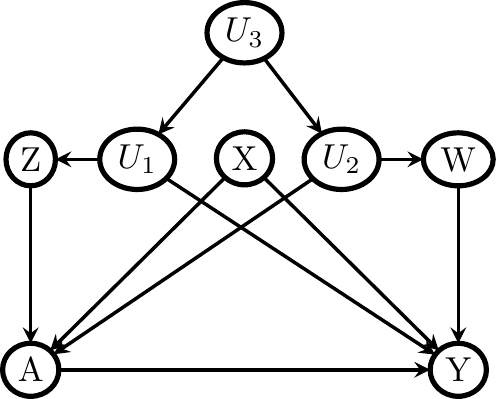}
		\subcaption{Coexistence of type (a)(b)(c) proxies when NUC fails.}
		\label{fig:dag1e}
	\end{minipage}
	\caption{Directed Acyclic Graphs illustrating treatment and outcome confounding proxies}
	\label{fig:dag1}
\end{figure}
Under assumptions $\left(  \ref{consistency}\right)  $-$\left(
\ref{positivity}\right)  ,$ it is well-known that
\begin{equation}
\beta\left(  a\right)  =%
{\displaystyle\sum\limits_{l}}
E\left(  Y|a,l\right)  f(l);\label{g-formula}%
\end{equation}
a formula most commonly known in the field of epidemiology as the
g-formula$^{1}$, a name associated with the work of James Robins which we
shall adopt in this paper.

It is interesting to consider alternative data generating mechanisms under
which assumption $\left(  \ref{exchangeability}\right)  $ holds, illustrated
in Figures 1.b. and 1.c, with the first of Type b where $L$ includes all
causes of $A$ that share an unmeasured common cause $U$ (and therefore are
associated) with $Y;$ while the second is of Type c where $L$ includes all
causes of $Y$ that share an unmeasured common cause $U$ (and therefore are
associated) with $A.$ These three possibilities may coexist, as displayed in
Figure 1.d. in which $L$ has been decomposed into three bucket types of
measured covariates $L=(X,W,Z),$ such that $X$ are measured covariates of Type
a, $Z$ are measured covariates of Type b, while $W$ are measured covariates of
Type c. At any rate, all settings represented in Figure 1 illustrate possible
data generating mechanisms under which exchangeability assumption $\left(
\ref{exchangeability}\right)  $ holds, without necessarily requiring that the
analyst identify which bucket type each covariate in $L$ belongs to.
Importantly, all settings given in Figure 1 rule out the presence of an
unmeasured common cause of $A$ and $Y,$ therefore ruling out unmeasured
confounding. Note that in order for exchangeability to hold in Figure 1.d, it
must be that as encoded in the DAG, unmeasured variables $U_{1}$ and $U_{2}$
are independent conditional on $A,X,Z$ and $W;$ otherwise, as illustrated in
Figure 1.e. the unblocked backdoor path $A-U_{2}-U_{3}-U_{1}-Y$ would
invalidate assumption $\left(  \ref{exchangeability}\right)  .$ As we show in
the next section, it is sometimes possible to relax this last assumption and
therefore exchangeability condition $\left(  \ref{exchangeability}\right)  $
while preserving identification of $\beta\left(  a\right)  $ despite the
presence of unmeasured confounding, provided that one can correctly identify
which bucket type each measured covariate falls into.

\section{\noindent Proximal identification in point exposure studies}
\begin{figure}[h]
	\centering
	\includegraphics[scale=1.1]{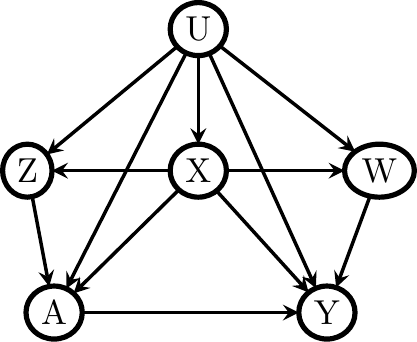}
	\caption{A DAG with endogenous point exposure and proxies}
	\label{fig:pointexpo}
\end{figure}
\noindent Next, consider Figure 2 which depicts a setting in which
exchangeability condition $\left(  \ref{exchangeability}\right)  $ fails,
despite having measured covariates $L=(X,Z,W),$ owing to the presence of an
unmeasured common cause $U$ of $A$ and $Y.$ This DAG may be viewed as
generalization of Figure 1.e by letting $U=(U_{1},U_{2},U_{3}).$ In the
following, we propose to replace the untestable assumption $\left(
\ref{exchangeability}\right)  $ with an assumption that the analyst has
correctly identified variables in bucket Types b and c. Formally, the causal
DAG Figure 2 implies that
\begin{align}
&  \left.  \left(  Y_{az},W_{az}\right)  \amalg\left(  A,Z\right)  \right\vert
U,X\label{NC point exposure}\\
&  \left.  Y_{az}=Y_{a}\right.  \label{NC point exposure 2}\\
&  \left.  W_{az}=W\right.  \label{NC point exposure 3}%
\end{align}

The first condition in the above display formally encodes the assumption that
adjusting for $(X,U)$ would in principle suffice to identify the joint causal
effect of $(A,Z)$ on $Y$ and $W$ respectively. This assumption is reasonable
as long as there exist a $U$ sufficiently enriched to include all common
causes of $(A,Z)$ and $\left(  Y,W\right)  $ not included in $X.$ As it is not
required that $U$ be observed, the assumption will generally hold even in
observational studies. The second assumption in the above display states that
$Z$ does not have a direct effect on $Y$ upon intervening on $A;$ likewise,
the third assumption states that $A$ and $Z$ do not have a causal effect on
$W.$ The first assumption will hold, provided that all variables in $Z$ are
correctly classified as of Type b, while the second assumption will hold
provided that variables in $W$ are correctly classified as of Type c. These
three assumptions imply that
\begin{align}
&  Y\amalg Z|A,U,X\label{NC.1}\\
&  W\amalg\left(  A,Z\right)  |U,X\label{NC.2}%
\end{align}

\noindent We formally refer to Type b variables $Z$ as treatment-inducing
confounding proxies and Type c variables $W$ as outcome-inducing confounding
proxies, provided that they satisfy assumptions $\left(  \ref{NC.1}\right)  $
and $\left(  \ref{NC.2}\right)  .$ It is important to note that $Z$ may not be
a direct cause of $A$ (in which case $Z\rightarrow A$ edge can be omitted in
Figure 2), and likewise $W$ may not be a direct cause of $Y$ (in which case
$W\rightarrow Y$ edge can be omitted in Figure 2), however as long as both are
$U$-relevant, and satisfy $\left(  \ref{NC.1}\right)  $ and $\left(
\ref{NC.2}\right)  ,$ they are considered valid proxies for our purposes and
may be allocated as type b or c at the analyst's discretion. We generally
favor assumptions $\left(  \ref{NC.1}\right)  $ and $\left(  \ref{NC.2}%
\right)  $ to $\left(  \ref{NC point exposure}\right)  $-$\left(
\ref{NC point exposure 3}\right)  $ as primitive identification conditions as
they do not require conceptualizing a potential intervention on covariates
$Z.$%

\noindent\textbf{Remark 1.} In prior work, variables of Types b and c
satisfying assumptions $\left(  \ref{NC.1}\right)  $ and $\left(
\ref{NC.2}\right)  $ have been called negative control exposure and negative
control outcome variables, referring to negative control variables an
investigator would need to supplement her observational study sample with,
such that $\left(  \ref{NC.1}\right)  $ and $\left(  \ref{NC.2}\right)  $
would be satisfied. In this paper, we prefer the proxy terminology to the
negative control nomenclature, to highlight the key observation, that often,
covariates measured in an observational study in an effort to control for
confounding, may not be sufficient to fulfil exchangeability, but nevertheless
can potentially be partitioned into proxies satisfying negative control
conditions $\left(  \ref{NC.1}\right)  $ and $\left(  \ref{NC.2}\right)  $.
This observation, therefore alleviates the need to supplement one's
observational study design by collecting additional data on potential negative
control variables, although variables of Type b and c may be enriched with
appropriately selected negative control auxiliary variables when available. \ 

\noindent\textbf{Remark 2. }It is further important to note that while we have
taken Figure 2 as canonical representation of proxy variables of Types b and
c, several alternative DAGs might be compatible with assumptions $\left(
\ref{NC.1}\right)  $ and $\left(  \ref{NC.2}\right)  ,$ as illustrated in the
Supplemental Appendix Table A.1.\ Interestingly, the DAG given in first row
and first column of Table A.1 of the appendix establishes that an instrumental
variable (IV) for the causal effect of $A$ on $Y$ may be included in Type a
bucket provided that it is also a valid IV\ for $W$. In fact, even an invalid
instrumental variable which fails to satisfy the IV independence
assumption$^{10,11}$ may also be included in bucket type b as $\left(
\ref{NC.1}\right)  $ and $\left(  \ref{NC.2}\right)  $ are satisfied$^{12}$. \ 

\noindent\textbf{Remark 3.} Additionally, one should note that similar to
exchangeability condition $\left(  \ref{exchangeability}\right)  $,
Assumptions $\left(  \ref{NC.1}\right)  $ and $\left(  \ref{NC.2}\right)  $
are not empirically testable as they presume certain null causal effects and
involve conditional independence statements given the unmeasured variable $U.$
Interestingly, the joint exclusion restriction $W_{za}=W$ is a given in
instances where $W$ and $Z$ are contemporaneous (and therefore cannot cause
each other), and as assumed throughout are pre-treatment covariates. \ The
treatment can therefore not have a causal effect on $W$ as the future cannot
cause the past. It is sometimes reasonable to include post-outcome variables
in $Z$ so that exclusion restriction conditions hold, again due to temporal
ordering. However, in order to satisfy conditions $\left(  \ref{NC.1}\right)
$ and $\left(  \ref{NC.2}\right)  $, there must be no unblocked causal pathway
between $\left(  Y,W\right)  $ and $Z$ conditional on $U,$ $X$ and $A.$
Likewise, it is sometimes possible to include pre-treatment measurements of
the outcome in view as potential negative control outcome and therefore to
include them in $W$, provided that they satisfy conditions $\left(
\ref{NC.1}\right)  $ and $\left(  \ref{NC.2}\right)  $ and therefore do not
have a direct effect on treatment and negative control exposure
variables$^{12-14}$. An example was provided in Miao and Tchetgen$\ $%
Tchetgen$^{12,13}$\ in context of studying the causal effect of air pollution
on say mortality or elderly hospitalization using time series data, in which
case air pollution measurement post-hospitalization\ may be a reasonable
choice of negative control exposure to include in $Z$, and hospitalization
measurement pre-air pollution may likewise be a reasonable negative control
outcome to include in $W.^{12,13}$

The aforementioned connection to negative control literature is instrumental
in determining sufficient conditions for nonparametric identification of
$\beta\left(  a\right)  $ by leveraging identification results recently
obtained by Miao et al.$^{15}$ We summarize their results below, provide
intuition for the results and refer the interested reader to their manuscript
for a careful treatment of mathematical conditions underpinning the approach.
In the next Section, we extend their results to the time-varying setting,
which to our knowledge is new to the literature, all proofs can be found in
the Appendix. \ 

Let $h\left(  a,x,w\right)  $ denote a solution to the equation:%
\begin{equation}
E(Y|a,z,x)=\sum_{w}h\left(  a,x,w\right)  f(w|a,x,z)\label{confounding bridge}%
\end{equation}
where in slight abuse of notation $\sum$ denotes an integral in case of
continuous $w.$ Next, suppose that the following conditions hold, for any
function $v\left(  \cdot\right)  $:%
\begin{equation}
E\left\{  v(U)|z,a,x\right\}  =0\text{ for all }z,a\text{ and }x\text{ if and
only if }v(U)=0;\label{completeness condition 1}%
\end{equation}
This condition is formally referred to as a completeness condition which
accommodates both categorical and continuous confounders. Completeness is a
technical condition taught in most foundational courses in theory of
statistical inference. Here one may interpret it as a requirement relating the
range of $U$ to that of $Z$ which essentially states that the set of proxies
must have sufficient variability relative to variability of $U.$ The condition
is easiest understood in the case of categorical $U,$ $Z$ and $W,$ with number
of categories $d_{u}$, $d_{z}$ and $d_{w}$ respectively$.$ In this case,
completeness requires that
\begin{equation}
min\left(  d_{z},d_{w}\right)  \geq d_{u}\label{categorical completeness}%
\end{equation}
which states that $Z$ and $W$ must each have at least as many categories as
$U.$ Intuitively, condition $\left(  \ref{categorical completeness}\right)  $
states that proximal causal learning can potentially account for unmeasured
confounding in the categorical case as long as the number of categories of $U$
is no larger than that of either proxies $Z$ and $W^{16}.$ This further
provides a rationale for measuring a rich set of baseline characteristics in
observational studies as a potential strategy for mitigating unmeasured
confounding via the proximal approach we now describe. Miao et al$^{15}$
established that under assumptions $\left(  \ref{NC.1}\right)  $-$\left(
\ref{completeness condition 1}\right)  ,$ the counterfactual mean
$\beta\left(  a\right)  $ can be identified nonparametrically by the formula:
\begin{equation}
\beta\left(  a\right)  =\sum_{w,x}h\left(  a,x,w\right)
f(w,x)\label{proximal g-formula}%
\end{equation}
We refer to equation $\left(  \ref{proximal g-formula}\right)  $ as the
\textit{proximal g-formula}, and to $h\left(  a,x,w\right)  $ as an outcome
confounding bridge function$^{15}$. A few key observations are in order.
First, equation $\left(  \ref{confounding bridge}\right)  $ defines a
so-called inverse problem formally known as a Fredholm integral equation of
the first kind. \ Formal conditions for existence of a solution of such an
equation are well established in functional analysis in mathematical
literature, but due to their technical nature are beyond the scope of the
current paper, though it is worth mentioning that existence of a solution
requires the following additional completeness condition:
\begin{equation}
E\left\{  v(Z)|w,a,x\right\}  =0\text{ for all }z,a\text{ and }x\text{ if and
only if }v(Z)=0;\label{completeness condition 2}%
\end{equation}
An assumption that cannot hold unless $W$ is $U$-relevant. \ We refer the
reader to Miao et al for a technical exposition of required regularity
conditions. In the categorical case, as established in Shi et al$^{16}$
condition $\left(  \ref{categorical completeness}\right)  $ along with a rank
condition for a certain matrix defined in terms of the conditional
distribution of $W$ given $\left(  Z,A,X\right)  $ suffices for equation
$\left(  \ref{confounding bridge}\right)  $ to admit a solution$^{16}$. It is
important to note that $h\left(  a,x,w\right)  $ satisfying $\left(
\ref{confounding bridge}\right)  $ need not be unique, any solution to this
equation yields the same value of the proximal g-formula. \ We also note that
by latent exchangeability, $\beta\left(  a\right)  =\sum_{u,x}E\left(
Y|a,u,x\right)  f(u,x)=\sum_{w,x}h\left(  a,x,w\right)  f(w,x);$ and as shown
in by Miao et al$^{15},$ $E\left(  Y_{a}|u,x\right)  =\sum_{w}h\left(
a,x,w\right)  f(w|u,x)$, which highlights the\ inverse-problem nature of the
task accomplished by proximal g-formula, which is to determine an $h$ that
satisfies this equality without explicitly modeling or estimating the latent
factor $U.$ A remarkable feature of proximal learning is that accounting for
$U$ without either measuring $U$ directly or estimating its distribution can
be accomplished provided that the set of proxies though imperfect, is
sufficiently rich so that the inverse-problem admits a solution in a
model-free framework.

Intuition about conditions under which a unique solution to equation $\left(
\ref{confounding bridge}\right)  $\ might exist can be gained in the simple
case of binary $A,W,Z,$ whereby it is straightforward to show that the unique
solution to $\left(  \ref{confounding bridge}\right)  $\ is
\begin{equation}
h(a,x,w)=E(Y|a,z,x)-g(a,x)\left[  w-\Pr\left(  W=1|a,z,x\right)  \right]
\label{explicit h}%
\end{equation}
where
\[
g(a,x)=\frac{E\left(  Y|a,z=1,x\right)  -E\left(  Y|a,z=0,x\right)  }%
{\Pr(W=1|a,z=1,x)-\Pr(W=1|a,z=0,x)}.
\]

Importantly, note that although the right-hand side to equation $\left(
\ref{explicit h}\right)  $\ appears to depend on~$z,$ the left-hand side
indicates that it does not, which is readily verified with some algebra. In
order for $h$ to be finite, one requires that $\Pr(W=1|a,z=1,x)-\Pr
(W=1|a,z=0,x)\neq0$; that is $W$ must be associated with $Z$ conditional on
$(A,X),$ a condition that one would expect to hold to the extent that $W$ and
$Z$ are strong proxies of $U,$ thus further highlighting the importance of
selecting strong potential proxies. In the binary case, $\beta\left(
a\right)  $ takes the closed form:
\[
\beta\left(  a\right)  =E_{X}\left\{  E(Y|a,Z,X)-g(a,X)\left[  \Pr\left(
W=1|a,X\right)  -\Pr\left(  W=1|a,Z,X\right)  \right]  \right\}
\]
A generalization of the above closed-form expression for proximal g-formula
with categorical variables is given in Shi et al$^{16}$ for the average causal
effect $\beta\left(  1\right)  -\beta\left(  0\right)  $ of a binary treatment
on the additive scale.\ Unfortunately, unlike the g-formula, the proximal
g-formula is not always available in closed-form and requires solving equation
$\left(  \ref{confounding bridge}\right)  $ numerically, which might be
computationally intensive and unstable due to its potential to be empirically
ill-posed. \ Ill-posedness in this case refers to the fact that small amount
of uncertainty in estimating the left handside of $\left(
\ref{confounding bridge}\right)  $ empirically can often induce excessive
uncertainty in obtaining a solution to the equation. \ Such ill-posedness is
typically addressed by some form of regularization of the integral equation.
Below, we describe a simple statistical modeling approach analogous to
g-computation, which sidesteps this difficulty by automatically generating
stable solutions to the equation under correct model specification. Revisiting
the motivating example given in the introduction, one may readily verify that
the structural equations $\left(  \ref{Linear Structural}\right)  $ imply that
there exists coefficient $\eta=\left(  \eta_{0},\eta_{a},\eta_{x}^{\prime
},\eta_{W}\right)  $ such that:%
\[
h(A,X,W;\eta)=\eta_{0}+\eta_{a}A+\eta_{x}^{\prime}X+\eta_{W}W
\]
and
\[
E\left(  Y-h(A,X,W;\eta)|A,Z,X\right)  =0
\]
so that equation $\left(  \ref{confounding bridge}\right)  $ is satisfied. By
then applying the proximal g-formula, one recovers $\beta_{a}=$ $\eta
_{a}=E\left\{  h(A=1,X,W;\eta)-h(A=0,X,W;\eta)\right\}  $ identifies the
causal effect parameter.

It is interesting to compare proximal g-formula to standard g-formula$^{1}$.
In this vein, suppose that $L=(X,W)$ suffices for exchangeability condition
$\left(  \ref{exchangeability}\right)  $, so that $Z=\varnothing$ ; then,
proximal g-formula reduces to the standard g-formula with $h\left(
a,x,w\right)  =E(Y|a,x,w)$ a stable solution to $\left(
\ref{confounding bridge}\right)  $ given that:%

\begin{align*}
E(Y|a,x) &  =\sum_{w}h\left(  a,x,w\right)  f(w|a,x)\\
&  =\sum_{w}E(Y|a,x,w)f(w|a,x);\\
\beta\left(  a\right)   &  =\sum_{w,x}h\left(  a,x,w\right)  f(w,x)\\
&  =\sum_{w,x}E(Y|a,x,w)f(w,x)
\end{align*}
From this perspective, exchangeability may be viewed as a form of
regularization of equation $\left(  \ref{confounding bridge}\right)  $ which
automatically yields a unique stable solution to the integral equation.

\noindent\textbf{Remark 4:}\noindent\ We note that Miao and Tchetgen
Tchetgen$^{12}$ considered alternative identifying conditions in that instead
of taking equation $\left(  \ref{confounding bridge}\right)  $ as starting
point, they a priori assume that there exist a bridge function $h(w,a,x)$ such
that $E(Y_{a}|u,x)=\sum_{w}h\left(  a,x,w\right)  f(w|u,x);$ in addition, they
replace completeness condition $\left(  \ref{completeness condition 1}\right)
$ which is not subject to an empirical test, with the testable completeness
condition that $E\left\{  v(w)|z,a,x\right\}  =0$ for all $z,a$ and $x$ if and
only if $v(w)=0;$ then they establish that such function $h$ must solve
equation $\left(  \ref{confounding bridge}\right)  .$

\section{Proximal identification in complex longitudinal studies}

We now consider proximal identification of causal effects in complex
longitudinal studies. In order to ground ideas and simplify the exposition, we
focus primarily on a special case of a longitudinal study with two follow-up
times and briefly review identification under a longitudinal version of
exchangeability. Thus, suppose that one has observed time-varying treatment
and covariate data $\left\{  L\left(  j\right)  ,A\left(  j\right)  \right\}
$ at follow up visits $j=0,1$ of a longitudinal study$.$ Let $Y$ denote the
outcome of interest measured at the end of follow-up $j=2$. We assume that
recorded data on the treatment and prognostic factors do not change except at
these times, moreover, $L(j)$ temporally precedes $A(j).$ We use overbars to
denote the history of that variable up to end of follow-up$;$ for example,
$\overline{L}=\left\{  L(0),L(1)\right\}  .$ \ Let $Y_{\overline{a}%
}=Y_{a(0),a(1)}$ denote the potential outcome had possibly contrary to fact, a
subject followed treatment regime $\overline{A}=\overline{a}.$ Our aim is to
identify the potential outcome mean $\beta\left(  \overline{a}\right)
=E\left(  Y_{\overline{a}}\right)  .$ To do so, three standard assumptions are
typically invoked. The first entails a longitudinal version of consistency:
\[
Y=Y_{\overline{A}}%
\]
linking counterfactual outcomes $\left\{  Y_{\overline{a}}:\overline
{a}\right\}  $ to observed variables $(Y,\overline{A}).$ \ The next assumption
is that there are no unmeasured confounders for the effect of $A(j)$ on $Y$,
that is, for all treatment histories $\overline{a,}$%

\begin{equation}
Y_{\overline{a}}\amalg A\left(  0\right)  |L\left(  0\right)  \text{ \ and
}Y_{\overline{a}}\amalg A\left(  1\right)  |A\left(  0\right)  =a\left(
0\right)  ,\overline{L}\label{sequential rand}%
\end{equation}
\ \ This assumption which generalizes exchangeability to the longitudinal
setting, is also known as the sequential randomization assumption (SRA)$^{17}%
$.\ It states that conditional on treatment history and the history of all
recorded covariates up to $j$, treatment at $j$ is essentially randomized by
nature and thus must be independent of the counterfactual random variable
$Y_{\overline{a}}.^{2,17}$ \ 

We finally assume that the following positivity assumption holds. \ For all
$a(j)$ in the support of $A(j)$%
\[
\text{if }f\left(  \overline{L}(j),\overline{A}(j-1)\right)  >0\text{ then
}f(a(j)|\overline{L}(j),\overline{A}(j-1))>0,\text{ }j=0,1
\]
with $A(-1)\equiv0,$ which essentially states that if any set of subjects at
time $j$ have the opportunity of continuing on a treatment regime
$\overline{a}$ under consideration, at least some will take that opportunity.
Robins established that under these assumptions, the counterfactual mean
$\beta\left(  \overline{a}\right)  $ is given by the longitudinal
g-formula$^{2,17}$:
\[
\beta\left(  \overline{a}\right)  =%
{\displaystyle\sum\limits_{\overline{l}}}
E\left(  Y|\overline{a},\overline{l}\right)
{\displaystyle\prod\limits_{j=1}^{2}}
f\left(  l(j)|\overline{l}(j-1),\overline{a}(j-1)\right)  .
\]

As argued in the introduction, in an observational study, the assumption of no
unmeasured confounding cannot be guaranteed to hold, and it is not subject to
empirical test, even when good efforts are made to collect data on crucial
covariates. As before, our aim is to relax sequential exchangeability/SRA by
explicitly incorporating measured covariates as proxies of underlying
confounding mechanisms longitudinally. \ In this vein, we suppose that one can
partition covariates measured at time $j=0,1$ into three bucket types
$L(j)=(Z(j),W(j),X(j)),j=0,1,$ such that the following conditions hold:
\begin{align}
&  \left(  \overline{Z},A(1)\right)  \amalg\left(  \overline{W},Y\left(
a_{1}\right)  \right)  |\overline{U},A(0),\overline{X}\label{long NC1}\\
&  \left(  Z(0),A(0)\right)  \amalg\left(  W(0),Y\left(  \overline{a}\right)
\right)  |U(0),X(0)\label{long NC2}%
\end{align}
These conditions are a longitudinal generalization of $\left(
\ref{NC point exposure}\right)  .$

\begin{figure}[h]
	\centering
	\includegraphics[scale=0.85]{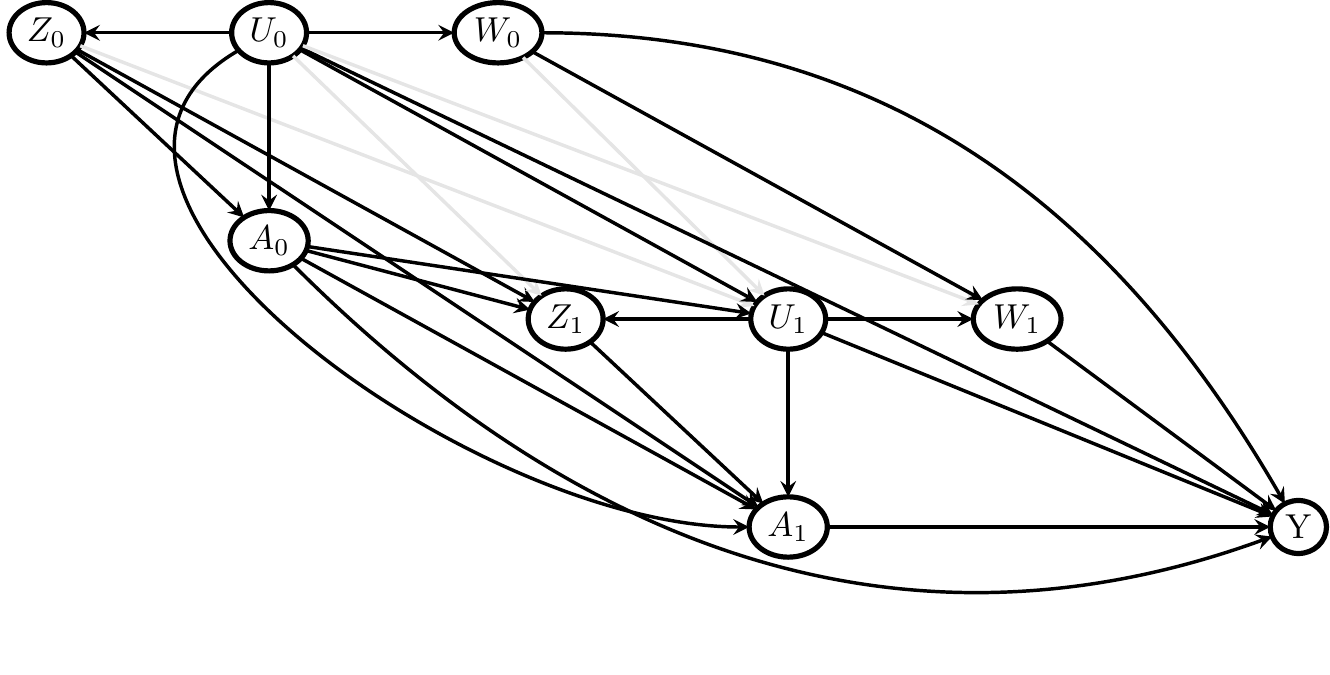}
	\caption{A DAG with endogenous time varying treatments and proxies}
	\label{fig:timedepexpo}
\end{figure}

Figure 3 illustrates a possible data generating mechanism in which conditions
$\left(  \pageref{long NC1}\right)  $ \ and $\left(  \ref{long NC2}\right)  $
hold, where to simplify the figure certain edges are shaded and we have
suppressed observed time-varying covariates $\overline{X}$ which structurally
follow the same relationship as $\overline{U}$ with other variables.
Additionally, for identification we require the following longitudinal
generalization of completeness condition $\left(
\ref{completeness condition 1}\right)  $, which state that for any function
$v\left(  \cdot\right)  :$
\begin{align}
E\left(  \nu\left(  \overline{U}\right)  |\overline{a},\overline{z}%
,\overline{x}\right)   &  =0\text{ if and only if }\nu\left(  \overline
{U}\right)  =0;\label{long completeness 1}\\
E\left(  \nu\left(  U\left(  0\right)  \right)  |a\left(  0\right)  ,z\left(
0\right)  ,x\left(  0\right)  \right)   &  =0\text{ if and only if }\nu\left(
U(0)\right)  =0.
\end{align}

Finally, extending assumption $\left(  \ref{confounding bridge}\right)  $ to
longitudinal setting, we suppose that there exist functions $H_{1}\left(
\overline{a}\right)  =h_{1}\left(  \overline{W},\overline{a},\overline
{X}\right)  $ and $H_{0}\left(  \overline{a}\right)  =h_{0}\left(  W\left(
0\right)  ,\overline{a},X\left(  0\right)  \right)  $ that solve equations:
\begin{align}
E\left(  Y|\overline{z},\overline{a},\overline{x}\right)   &  =E\left(
H_{1}\left(  \overline{a}\right)  |\overline{z},\overline{a},\overline
{x}\right)  \label{confounding bridge long1}\\
E\left(  H_{1}\left(  \overline{a}\right)  |z\left(  0\right)  ,\overline
{a},x\left(  0\right)  \right)   &  =E\left(  H_{0}\left(  \overline
{a}\right)  |z\left(  0\right)  ,\overline{a},x\left(  0\right)  \right)
\label{confounding bridge long2}%
\end{align}
Confounding bridge equation $\left(  \ref{confounding bridge long1}\right)  $
is exactly equivalent to equation $\left(  \ref{confounding bridge}\right)  $
with\ $\left(  \overline{W},\overline{Z},\left\{  \overline{X},A_{0}\right\}
\right)  $ replacing $\left(  W,Z,X\right)  ,$ and $A(1)$ replacing $A.$ As
show in the result below, this assumption yields identification of $E\left(
Y_{a(1)}|a(0),\overline{x}\right)  $, while the second assumption $\left(
\ref{confounding bridge long2}\right)  $ which does not have a point exposure
analog, yields identification of $E\left(  Y_{\overline{a}}|x(0)\right)  $. In
fact, we have the following result.

\underline{Result 1}:Suppose that assumptions $\left(  \ref{long NC1}\right)
$-$\left(  \ref{confounding bridge long2}\right)  $\ are satisfied, then we
have that%
\begin{align*}
E\left(  Y_{a(1)}|a(0),\overline{u},\overline{x}\right)   &  =E\left(
H_{1}\left(  \overline{a}\right)  |a(0),\overline{u},\overline{x}\right)  \\
E\left(  Y_{\overline{a}}|\ u\left(  0\right)  ,x\left(  0\right)  \right)
&  =E\left\{  H_{0}\left(  \overline{a}\right)  |u\left(  0\right)
,x(0)\right\}
\end{align*}
and
\begin{align*}
E\left(  Y_{a(1)}|a(0),\overline{x}\right)   &  =E\left(  H_{1}\left(
\overline{a}\right)  |a(0),\overline{x}\right)  \\
E\left(  Y_{\overline{a}}|x(0)\right)   &  =E\left\{  H_{0}\left(
\overline{a}\right)  |x(0)\right\}  \\
\beta\left(  \overline{a}\right)   &  =E\left(  Y_{\overline{a}}\right)
=E\left\{  H_{0}\left(  \overline{a}\right)  \right\}  =E\left\{  h_{0}\left(
W\left(  0\right)  ,\overline{a},X\left(  0\right)  \right)  \right\}
\end{align*}

Result 1 can be extended to a longitudinal study of follow-up length $J>2$ as
shown in the Appendix; the proof of which implies Result 1 as a special case.
As in the point treatment setting, $H_{1}\left(  \overline{a}\right)  $ and
$H_{0}\left(  \overline{a}\right)  $ need not be uniquely identified in order
for $\beta\left(  \overline{a}\right)  $ to be uniquely identified.

\section{Proximal g-computation}

\noindent In this section we describe a practical approach for estimating the
proximal g-formula. we first describe the approach in the point treatment case
before extending it to the case of time-varying treatment. Thus, suppose that
one has observed an i.i.d sample of size $n$ on $(A,L=\left(  X,Z,W\right)
).$ It is then convenient to directly specify a parametric model for the
outcome bridge function:%
\[
h\left(  W,A,X\right)  =h\left(  W,A,X;\eta\right)  ,
\]
with unknown parameter $\eta;$ and for the joint law
\[
f\left(  L,A\right)  =f\left(  L,A;\theta\right)  ;\text{ }%
\]
with unknown parameter $\theta.$ Note that together, these models entail a
parametric model for
\[
\mu\left(  A;\eta,\theta\right)  =E\left(  Y|A,X,Z;\eta,\theta\right)  =%
{\displaystyle\sum\limits_{w}}
f(w|Z,A,X;\theta)h\left(  w,A,X;\eta\right)
\]
in terms of $\theta$ and $\eta.$ This modeling assumption is therefore
appropriate only if the outcome mean admits the representation given above. As
we show below, directly modeling the outcome confounding bridge function
obviates the need to solve complicated integral equations which are well-known
to be ill-posed and therefore to admit unstable solutions. The above modeling
strategy can be viewed as a form of regularization of the problem so as to
resolve ill-posedness. Although not pursued here, a variety of semiparametric
(e.g. partially linear model, single index model) or nonparametric (e.g.
generalized additive, reproducing kernels, neural networks) may be used to
model $h$ more flexibly, thus alleviating concerns about specification bias.

Let $\widehat{\theta}$ denote the maximum likelihood estimator of $\theta,$
and define $\widehat{f}_{W|Z,A}(W)=f\left(  W|Z,A,X;\widehat{\theta}\right)  $
implied by $f\left(  L,A;\widehat{\theta}\right)  .$ One may then estimate
$\eta$ based on Result 1, by fitting via least-squares, the regression model:
$\mu\left(  A,\eta\right)  =E\left(  Y|A,X,Z\right)  =E\left(  H\left(
\eta\right)  |A,X,Z\right)  $ given by
\begin{equation}
\widehat{\mu}\left(  A;\eta\right)  =%
{\displaystyle\sum\limits_{w}}
\widehat{f}_{W|Z,A}(w)h\left(  w,A,X;\eta\right)  .\label{surrogate outcome}%
\end{equation}
For continuous $Y,$ this may be accomplished by least-squares minimization:%
\[
\ \widehat{\eta}=\arg\min_{\eta}E_{n}\left\{  Y-\widehat{\mu}\left(
A;\eta\right)  \right\}  ^{2}%
\]
where $E_{n}$ stands for sample average. \ Then, assuming all models are
correctly specified, one can show that $\widehat{\beta}\left(  a\right)  $ is
a consistent and asymptotically normal estimator of $\beta\left(  a\right)  $,
where
\[
\widehat{\beta}\left(  a\right)  =E_{n}\left\{  h\left(  W,a,X;\widehat{\eta
}\right)  \right\}  .
\]
For inference, we recommend using the nonparametric bootstrap to obtain
standard errors and confidence intervals. \ We note that evaluating $\left(
\ref{surrogate outcome}\right)  $ might require evaluating either a sum, an
integral or both with respect to a high dimensional variable $w;$ in many
cases, the sum/integral may not admit a closed form expression or may be
computationally prohibitive to evaluate, in which case Monte Carlo
approximation of $\widehat{\mu}\left(  A,\eta\right)  $ may provide a
practical solution. In case of binary $Y$, use of a link function (say logit
or probit link function) may be necessary in specifying model for $h\left(
w,A,X;\eta\right)  $, in order to ensure that $\mu\left(  A;\eta
,\theta\right)  =\Pr\left(  Y=1|A,Z,X;\eta,\theta\right)  $ lies in the unit
interval (0,1). Estimation in the binary case can then proceed by standard
maximum likelihood estimation thus maximizing the log likelihood function
\[
\ \widehat{\eta}=\arg\max_{\eta}E_{n}\left\{  Y\log\widehat{\mu}\left(
A;\eta\right)  +\left(  1-Y\right)  \log\left(  1-\widehat{\mu}\left(
A;\eta\right)  \right)  \right\}
\]
In the Appendix, we describe special cases where $\widehat{\mu}\left(
A,\eta\right)  $ admits a closed form expression. Here, we discuss the
important special case of proximal g-computation under a linear specification
for $h\left(  W,A,X;\eta\right)  ,$ say
\[
h\left(  W,A,X;\eta\right)  =\beta_{a}A+\eta_{w}^{\prime}W+\eta_{x}^{\prime}X;
\]
so that $E(Y_{a})=\beta\left(  a\right)  =\beta_{0}+\beta_{a}a$ where
$\beta_{0}=\eta_{w}^{\prime}E\left\{  W\right\}  +\eta_{x}^{\prime}E\left\{
X\right\}  $ where $\eta_{x}^{\prime}X$ in includes an intercept term.
\ Suppose further that one specifies a (multivariate) linear regression model
\begin{equation}
W=\left(  1,Z^{\prime},A,X^{\prime}\right)  \Theta+\varepsilon_{W}%
\label{Wmodel}%
\end{equation}
Then, one can estimate the average causal effect $\beta_{a}=E(Y_{a+1}-Y_{a})$
with the regression coefficient $\widehat{\beta}_{a}$ obtained by fitting the
standard linear regression model:%
\begin{equation}
Y=\beta_{a}A+\eta_{w}^{\prime}\widehat{W}+\eta_{x}^{\prime}X+\varepsilon
_{y}\label{second stage}%
\end{equation}
by least-squares, where $\widehat{W}=\left(  1,Z^{\prime},A,X^{\prime}\right)
\widehat{\Theta}$ is the element-wise least-squares regression of $W$ on
$\left(  1,Z^{\prime},A,X^{\prime}\right)  ^{12}.$ We refer to this procedure
as proximal two-stage least squares (P2SLS) given its close relationship to
2SLS\ estimation in instrumental variable setting$^{18}$. This connection in
fact has implications for practice as it indicates that the estimator can be
implemented with any off-the-shelf instrumental variable software which can
perform 2SLS for multivariate exposure variable upon taking $W$ as the
endogenous (multivariate)\ variable, $Z$ playing the role of IV, with $A$
taken as a covariate. Such software can be used to obtain $\widehat{\beta}%
_{a}$ and corresponding confidence intervals, accounting for the uncertainty
in the first stage estimator $\widehat{W}.$ \ The model given in equation
$\left(  \ref{second stage}\right)  $ has an interesting interpretation as it
emulates a standard regression adjustment of confounding by $X,$ and further
adjusts for $\widehat{W}$ as proxy for the unmeasured factor $U$ therefore
deconfounding the standard regression approach. As mentioned in the
introduction, we can therefore refer to $\widehat{W}$ as \textit{proximal
control variable}.

\noindent\textbf{Remark 5. }One may note from the description of P2SLS that in
the event that dim$\left(  Z\right)  $%
$<$%
dim$\left(  W\right)  ,$ $\eta_{w}$ in $\left(  \ref{second stage}\right)  $
may not be uniquely identified; nonetheless, it is straightforward to verify
that all least squares solutions for $\widehat{\eta}_{w}$ yield a consistent
estimator $\widehat{\beta}_{a}.$ Either way, a test of presence of confounding
bias entails a standard statistical test of the null hypothesis that all
components of $\eta_{w}$ are identically zero, which is readily available even
when dim$\left(  Z\right)  $%
$<$%
dim$\left(  W\right)  .$ 

Next, we consider the longitudinal setting where we observe an i.i.d sample of
size $n$ on $(\overline{A},\overline{L}=\left(  \overline{X},\overline
{Z},\overline{W}\right)  )$. Then, parametric proximal g-computation relies on
specifying parametric models for the outcome bridge functions:%
\begin{align*}
h_{1}\left(  \overline{W},\overline{A},\overline{X}\right)   &  =h_{1}\left(
\overline{W},\overline{A},\overline{X}\left(  j\right)  ;\eta_{1}\right)  ,\\
h_{0}\left(  W(0),\overline{A},X(0)\right)   &  =h_{0}\left(  W(0),\overline
{A},X(0);\eta_{0}\right)
\end{align*}
with unknown parameter $\eta_{j};$ and for the joint law
\[
f\left(  \overline{L},\overline{A}\right)  =f\left(  \overline{L},\overline
{A};\theta\right)  ;\text{ }%
\]
with unknown parameter $\theta.$ Let $\widehat{\theta}$ denote the maximum
likelihood estimator of $\theta,$ and define $\widehat{f}_{1}(\overline
{W})=f\left(  \overline{W}|\overline{Z},\overline{A},\overline{X}%
;\widehat{\theta}\right)  $ and $\widehat{f}_{0}(W\left(  0\right)  )=f\left(
w(0)|A(0),X(0),Z(0);\widehat{\theta}\right)  $ both deduced from $f\left(
\overline{L},\overline{A};\widehat{\theta}\right)  .$ Then we propose to
estimate $\eta_{j}$ based on Result 1, by recursively fitting regression
models of $Y$ on $\mu_{1}\left(  \overline{A};\theta,\eta_{1}\right)
=E\left(  H_{1}\left(  \eta_{1}\right)  |\overline{A},\overline{X}%
,\overline{Z};\theta\right)  ,$ and of $H_{1}\left(  \eta_{1}\right)  $ on
$\mu_{0}\left(  \overline{A};\eta_{0},\theta\right)  =E\left(  H_{0}\left(
a_{1},\eta_{0}\right)  |A(0),X(0),Z(0);\theta\right)  $ given by
\begin{align*}
\widehat{\mu}_{1}\left(  \eta_{1}\right)   &  =%
{\displaystyle\sum\limits_{\overline{w}}}
\widehat{f}_{1}(\overline{w})h_{1}\left(  \overline{w},\overline{A}%
,\overline{X};\eta_{1}\right)  ,\\
\widehat{\mu}_{0}\left(  A(1);\eta_{0}\right)   &  =%
{\displaystyle\sum\limits_{w(0)}}
\widehat{f}_{0}(w\left(  0\right)  )h_{0}\left(  w\left(  0\right)
,\overline{A},X(0);\eta_{0}\right)
\end{align*}
Each of these regressions can readily be performed via ordinary least-squares.
Then, assuming all models are correctly specified and Result 1 holds, one can
show that $\widehat{\beta}\left(  \overline{a}\right)  $ is a consistent
estimator of $\beta\left(  \overline{a}\right)  $, and is approximately
normally distributed, where
\[
\widehat{\beta}\left(  \overline{a}\right)  =E_{n}\left\{  h_{0}\left(
W\left(  0\right)  ,\overline{a},X\left(  0\right)  ;\widehat{\eta}%
_{0}\right)  \right\}  .
\]
In order to estimate standard errors for $\widehat{\beta}\left(  \overline
{a}\right)  $ and confidence intervals for $\beta\left(  \overline{a}\right)
$, we recommend using the nonparametric bootstrap$^{19}$. We refer to the
above estimation procedure as parametric proximal g-computation, the proximal
analog to parametric g-computation algorithm of Robins$^{2,17}$.

The algorithm simplifies tremendously in case of additive confounding bridge
functions, say:%
\begin{align}
h_{1}\left(  \overline{W},\overline{A},\overline{X}\left(  j\right)  ;\eta
_{1}\right)   &  =\left(  1,\mathrm{c}_{a}\left(  \overline{A}\right)
,\mathrm{c}_{w}\left(  \overline{W}\right)  ^{\prime},\mathrm{c}_{w}\left(
\overline{X}\right)  ^{\prime},X^{\prime}(0)\right)  \eta_{1}%
,\label{Linear h1}\\
h_{0}\left(  W(0),\overline{A},X(0);\eta_{0}\right)   &  =\left(
1,\mathrm{c}_{a}\left(  \overline{A}\right)  ,W^{\prime}\left(  0\right)
,X^{\prime}(0)\right)  \eta_{0}, \label{linear h0}%
\end{align}
where for time-varying variable $B(j),\mathrm{c}_{b}\left(  \overline
{B}\right)  $ denotes a user specified function of $\overline{B},$ for
instance, we might take $\mathrm{c}_{b}\left(  \overline{B}\right)
=\mathrm{cum}\left(  \overline{B}\right)  =B(0)+B(1),$ where in case of vector
$B$, the sum applies entry-wise such that $\mathrm{c}_{b}\left(  \overline
{B}\right)  $ is a vector of the same dimension as $B(j)$. \ Then proximal
g-computation can be implemented by the following recursive least-squares algorithm:

\underline{Proximal recursive least squares algorithm:}

\noindent Step 1: fit the multivariate linear regression
\begin{equation}
\mathrm{c}_{w}\left(  \overline{W}\right)  =\left(  1,\mathrm{c}_{z}\left(
\overline{Z}\right)  ^{\prime},\mathrm{c}_{a}\left(  \overline{A}\right)
^{\prime},\mathrm{c}_{x}\left(  \overline{X}\right)  ^{\prime},X^{\prime
}(0)\right)  \Theta_{1}+\mathbf{\varepsilon}_{W} \label{cumW model}%
\end{equation}
by applying least-squares separately to each entry of vector $\mathrm{c}%
\left(  \overline{W}\right)  $, and let
\[
\widehat{\mathrm{c}}_{w}=\left(  1,\mathrm{c}_{z}\left(  \overline{Z}\right)
^{\prime},\mathrm{c}_{a}\left(  \overline{A}\right)  ^{\prime},\mathrm{c}%
_{x}\left(  \overline{X}\right)  ^{\prime},X^{\prime}(0)\right)
\widehat{\Theta}_{1}%
\]
denote its fitted values;

\noindent Step 2:\ fit the linear regression
\[
Y=\left(  1,\mathrm{c}_{a}\left(  \overline{A}\right)  ^{\prime}%
,\widehat{\mathrm{c}}_{w}^{\prime},\mathrm{c}_{x}\left(  \overline{X}\right)
^{\prime},X^{\prime}(0)\right)  \eta_{1}+\mathbf{\varepsilon}_{Y}%
\]
by least-squares where we note that $\widehat{\mathrm{c}}_{w}\left(
\overline{W}\right)  $ has been substituted in for $\mathrm{c}\left(
\overline{W}\right)  ,$ and let
\[
\widehat{H}_{1}\left(  \overline{A}\right)  =\left(  1,\mathrm{c}_{a}\left(
\overline{A}\right)  ^{\prime},\mathrm{c}_{w}\left(  \overline{W}\right)
^{\prime},\mathrm{c}_{x}\left(  \overline{X}\right)  ^{\prime},X^{\prime
}(0)\right)  \widehat{\eta}_{1}%
\]

\noindent Step 3: fit the multivariate linear regression
\begin{equation}
W(0)=\left(  1,Z(0)^{\prime},\mathrm{c}_{a}\left(  \overline{A}\right)
^{\prime},X(0)^{\prime}\right)  \Theta_{0}+\mathbf{\varepsilon}_{W}
\label{cumW0 model}%
\end{equation}
by applying least-squares separately to each entry of vector $W(0)$, and let
\[
\widehat{W}(0)=\left(  1,Z(0)^{\prime},\mathrm{c}_{a}\left(  \overline
{A}\right)  ^{\prime},X(0)^{\prime}\right)  \widehat{\Theta}_{0}%
\]
denote its fitted values; fit the linear regression%
\[
\widehat{H}_{1}=\left(  1,\mathrm{c}_{a}\left(  \overline{A}\right)  ^{\prime
},\widehat{W}^{\prime}\left(  0\right)  ,X^{\prime}(0)\right)  \eta
_{0}+\varepsilon_{h_{1}}%
\]
by least-squares, to obtain an estimate of $H_{0}$,
\[
\widehat{H}_{0}\left(  \overline{A}\right)  =\left(  1,\mathrm{c}_{a}\left(
\overline{A}\right)  ^{\prime},W^{\prime}\left(  0\right)  ,X^{\prime
}(0)\right)  \widehat{\eta}_{0};
\]

\noindent Step 4:Evaluate%
\[
\widehat{\beta}\left(  \overline{a}\right)  =E_{n}\left\{  \widehat{H}%
_{0}\left(  \overline{a}\right)  \right\}  =\left(  1,\mathrm{c}_{a}\left(
\overline{a}\right)  ^{\prime},E_{n}\left\{  W^{\prime}\left(  0\right)
\right\}  ,E_{n}\left\{  X^{\prime}(0)\right\}  \right)  \widehat{\eta}_{0}.
\]

It is important to note that although additive, the specific form of models
used in Steps 1-4 can be quite flexible and can accommodate both
nonlinearities (e.g. using either polynomial or splines to model covariates)
as well as interactions with treatment or among covariates. More flexible
models can be somewhat more involved as each nonlinear specification of
$\overline{W}$ entries requires a corresponding regression in Step 1. We also
note that the particular manner in which a treatment or covariate history
enters a given model is entirely to the discretion of the analyst. For
instance, natural options for $\mathrm{c}_{a}\left(  \overline{A}\right)  $
include $(A(0),A(1),A(0)\times A(1)),\mathrm{cum}\left(  \overline{A}\right)
$ or simply $A(1)$ as viable alternatives depending on their respective goodness-of-fit.

Interestingly, under linearity of $H_{0}$ and $H_{1}$ with respect to
$\overline{W},$ the proximal recursive least squares algorithm yields an
estimator of $\beta\left(  \overline{a}\right)  ,$ that remains consistent
even if linear models $\left(  \ref{cumW model}\right)  $ and $\left(
\ref{cumW0 model}\right)  $ for $\overline{W}$ and $W(0)$ are misspecified.
Likewise, in the point treatment setting, $\widehat{\beta}\left(  a\right)  $
can be shown to remain consistent even if model $\left(  \ref{Wmodel}\right)
$ is not correctly specified provided that $h_{0}$ and $h_{1}$ are correctly
specified. The implication of this result is that OLS provides extra
protection against model misspecification bias in modeling $\overline{W}$ as a
linear model, including for binary or discrete components of $W.$ This is an
important property that does not generally hold for proximal g-computation
algorithm which requires in addition to correct specification of $h_{0}$ and
$h_{1}$, that one also specify $f\left(  \overline{L},\overline{A}%
;\theta\right)  $ correctly. It is however possible to obtain an estimator of
$h_{0}\left(  \eta_{0}\right)  $ and $h_{1}\left(  \eta_{1}\right)  $ using a
recursive generalized methods of moments (RGMM) which does not require a model
for $f\left(  \overline{L},\overline{A}\right)  $ and therefore is not
susceptible to bias due to modeling the latter incorrectly$.$ We refer the
interested reader to Miao and Tchetgen Tchetgen$^{12}$ in point treatment
case. A detailed treatment of this more robust estimation approach in
longitudinal settings will be described elsewhere. It is worth noting that
when $f\left(  \overline{L},\overline{A};\theta\right)  $ is correctly
specified, one can generally expect proximal g-computation to be more
efficient than proximal recursive least square and recursive generalized
method of moments.

\section{Data applications}

\subsection{Point treatment application}

We first illustrate proximal estimation of causal effects in a point treatment
application to the Study to Understand Prognoses and Preferences for Outcomes
and Risks of Treatments (SUPPORT) with the aim of evaluating the causal effect
of right heart catheterization (RHC) during the initial care of critically ill
patients in the intensive care unit (ICU) on survival time up to 30
days$^{20}$. RHC was performed in 2184 patients within the initial 24 hours of
ICU stay, while 3551 patients were managed without RHC. The SUPPORT study
collected rich patient information encoded in 73 covariates, including
demographics (such as age, sex, race, education, income, and insurance
status), estimated probability of survival, comorbidity, vital signs,
physiological status, and functional status. The outcome of interest is the
number of days between admission and death or censoring at 30 days ($Y$). Ten
variables measuring the patient's overall physiological status were measured
from a blood test during the initial 24 hours in the ICU: \texttt{serum
sodium, serum potassium, serum creatinine, bilirubin, albumin, PaO2/(.01*FiO2)
ratio, PaCO2, serum PH (arterial), white blood cell count}, and
\texttt{hematocrit}. These variables may be subject to substantial measurement
error and as single snapshot of underlying physiological state over time may
be viewed as potential confounding proxies. Among the ten physiological status
measures, four \texttt{(pafi1, paco21, ph1, hema1)} are strongly correlated
with both the treatment and the outcome; thus we construct proxies $Z$ and $W$
from this reduced set of variables and collect all 67 remaining variables as
covariates ($X$).

As we hypothesize that our four candidate proxies are equally likely to be
valid treatment-inducing proxy or outcome-inducing proxy, we consider a
practical strategy for allocating the four candidate proxies to bucket types b
and c. The approach first ranks proxies according to their strength of
association in treatment (based on logistic regression of $A$ on $L$) and
outcome models (based on linear regression of $Y$ given $A$ and $L)$
respectively; next, we select proxies in decreasing order of strength of
association, first selecting the proxy with strongest association with the
outcome as outcome-inducing proxy and likewise for the treatment. In case of a
tie, that is if these are the same variable, one may either decide to
prioritize one of the two buckets or alternatively to randomize allocation to
a proxy bucket type; upon allocating a given variable, say to the
outcome-inducing proxy bucket, one subsequently removes the variable from the
list of remaining treatment-inducing candidate proxies, and vice-versa. The
algorithm stops when all proxies have been allocated. The algorithm produced
the allocation $Z=(\mathtt{pafi1,paco21})$ and $W=(\mathtt{ph1,hema1}).$ For
estimation of the causal effect of interest, we assume a linear outcome
confounding bridge function,
\[
h(a,x,w)=\eta_{0}+\eta_{a}a+\eta_{x}^{\prime}x+\eta_{w}^{\prime}w,
\]
in which case, the coefficient $\eta_{a}=\beta_{a}=E\left\{
h(1,X,W)-h(0,X,W)\right\}  $ encodes the causal effect of interest. After
allocating the proxies, we apply two stage least squares to estimate the
confounding bridge function, which can be implemented via routine R software
such as \texttt{ivreg}. The following is a standard call of \texttt{ivreg} in R,%

\[
\mathrm{ivreg(Y\sim A+X+W|A+X+Z,data=rhc)}%
\]
Ordinary least squares results in a negative and statistically significant
causal effect estimate $\widehat{\beta}_{a}\left(  \text{OLS}\right)  =$
$-1.25$ with standard error=$0.28$. Outcome-inducing proxy $\mathtt{ph1}$ is
associated with confounding bridge parameter $\left(  \widehat{\eta}%
_{w}=-16.92,\text{ standard error=}8.8\right)  $, indicating moderate
empirical evidence that unmeasured confounding might be biasing
$\widehat{\beta}_{a}\left(  \text{OLS}\right)  $. The causal effect estimate
obtained by P2SLS is substantially larger than standard OLS point estimate
$\widehat{\beta}_{a}\left(  \text{Proximal}\right)  =-1.80$ with corresponding
standard error= $0.43$. These results suggest that RHC may have an even more
harmful effect on 30 day-survival among critically ill patients admitted into
an ICU than previously documented. Results from this analysis are summarized
in tables provided in the Supplemental Appendix. 

\subsection{Time-varying treatment application}

\noindent We reanalyze data from an article published by Choi et al
(2002)$^{21}$\ on the potential protective effects of the anti-rheumatic
therapy Methotrexate (MTX) among patients with rheumatoid arthritis. While
Choi et al focused on survival as an endpoint and used a Cox marginal
structural models to quantify joint treatment effects under SRA, here we
consider the joint causal effects of MTX on average of reported number of
tender joints, an important measure of disease progression, without appealing
to SRA. Our analysis includes individuals who were older than age 18 years and
who attended the Wichita Arthritis Center at least twice between Jan 1, 1981
(when weekly low-dose methotrexate therapy and health assessment questionnaire
scores became available) and Dec 31, 1999; had rheumatoid arthritis fulfilling
the 1958-1987 American College of Rheumatology (formerly the American
Rheumatism Association) criteria for rheumatoid arthritis; and had not
received methotrexate before their first visit to the center, who survived
more than 12 months.

Methotrexate use and dose was recorded in the computer database at each clinic
visit. We classified methotrexate exposure status as ever-treated or
never-treated, i.e., once a patient starts methotrexate therapy, he or she was
considered on therapy for the rest of the follow-up. This approach provides a
conservative estimate of methotrexate efficacy just as intent-to-treat
analysis does in randomized clinical trial.

A thousand and ten patients with rheumatoid arthritis met our inclusion
criteria, 183 of them were treated with methotrexate at month 6 of follow-up.
We have recorded baseline covariates including age, sex, past smoking status,
education level, rheumatoid arthritis duration, calendar year and rheumatoid
factor positive. Time varying covariates include current smoking status,
health assessment questionnaire, number of tender joints, patient's global
assessment, erythrocyte sedimentation rate, number of disease modifying
antirheumatic drugs taken and prednisone use. Our objective is therefore to
evaluate the joint effects of MTX use at baseline and month six on average of
tender joints at month 12 of follow-up. In addition to proximal learning, for
comparison, similar to Choi et al, we also evaluated the causal effect of
interest under a marginal structural linear model $E(Y_{\overline{a}}%
)=\beta_{0}+\beta_{a}\mathrm{cum}\left(  \overline{a}\right)  =\beta_{0}%
+\beta_{a}\left\{  a\left(  0\right)  +a(1)\right\}  \mathrm{\ }$where $a(0)$
and $a(1)$ are MTX use at baseline and at month 6 respectively, estimated via
standard inverse probability weighted least squares assuming SRA given both
all baseline and time-varying covariates. \ 

We then implemented proximal recursive least squares algorithm under linear
outcome confounding bridge specification $\left(  \ref{Linear h1}\right)  $
and $\left(  \ref{linear h0}\right)  ,$ with $X=$(age, education, sex,
smoking, rheumatoid arthritis duration, calendar year). Since number of tender
joints at one year of follow-up is the primary outcome, tender joints count
(\textrm{jc)} at baseline and at follow-up month 6 are both natural candidate
as outcome-inducing proxies. Other candidate proxies included health
assessment questionnaire (haqc), patient's global assessment of disease status
(gsc) and erythrocyte sedimentation rate (esrc), number of disease-modifying
antirheumatic drugs (dmrd), rheumatoid factor positive (rapos) and prednisone
use (onprd2). We further reduced the set of candidate proxies to candidate
variables associated with both treatment and outcome variables. Finally, we
applied the allocation algorithm described in the prior section resulting in
$Z\left(  j\right)  =\mathrm{haqc}\left(  j\right)  $ and $W\left(  j\right)
=\mathrm{jc(j)}.$

IPW least squares suggests a protective effect of MTX with $\widehat{\beta
}_{\overline{a}}=$-0.23 (-0.43, -0.02), although validity of this finding is
contingent on SRA. Proximal recursive least-squares yields results suggests a
stronger protective effect $\widehat{\beta}_{\overline{a}}=$-0.37 (-0.67,
-0.13), with strong evidence of confounding bias $\left(  \widehat{\eta}%
_{w,0},\widehat{\eta}_{w,1}\right)  =(0.785,0.524)$ with corresponding 95\%
confidence intervals $(0.50,1.1)$ and $(0.33,0.71)$ respectively. These
results reinforce understanding of potential protective effects of MTX on
disease progression. Results from this analysis are summarized in tables
provided in the Supplemental Appendix. 

\section{Discussion}

We have described a new framework for the analysis of observational data
subject to potential confounding bias. The approach acknowledges that in
practice, measured covariates generally fail in observational settings to
capture all potential confounding mechanisms and at most may be seen as proxy
measurements of underlying confounding factors. \ Our proximal causal learning
framework provides a formal potential outcome framework under which one can
articulate conditions to identify causal effects from proxies. \ We have
described proximal g-formula and proximal g-computation algorithm for
estimation in point treatment and time-varying treatment settings. \ The
proximal approach is closely related to negative control methods recently
proposed for detection and sometimes estimation of point treatment
interventions$^{12,13,16,22}$. We refer the reader to Shi et al$^{23}$ for a
recent review of negative control literature.

While similar to standard g-computation, our proximal g-computation algorithm
(as well as proximal two stage least squares and recursive least
squares)\ rely on correct specification of outcome confounding bridge
functions, we are currently in the process of developing alternative methods
which similar to inverse-probability weighting, rely on a model for a
so-called treatment confounding bridge function such that it is possible to
construct two separate estimators of the average treatment effect each
depending on a different model; either outcome or treatment confounding bridge
function. Interestingly, we have also developed doubly robust estimators that,
similar to standard doubly robust estimators developed by Robins and
colleagues$^{24}$, remain unbiased in large samples provided at least one
confounding bridge function model is correct, but not necessarily both.
\ These results along with further evaluation of finite sample performance of
proximal inference will be presented in future papers.

\begin{center}
\bigskip{\Large Appendix }
\end{center}

\underline{\noindent Proof of Result 1: }We establish the following general
result which implies Result 1:suppose that for $j=J-1,....,0$%
\[
\left(  \overline{Z}\left(  j\right)  ,A\left(  j\right)  \right)
\amalg\left(  \overline{W}\left(  j\right)  ,Y\left(  \overline{a}\right)
\right)  |\overline{U}\left(  j\right)  ,\overline{A}\left(  j-1\right)
=\overline{a}(j-1),\overline{X}\left(  j\right)  ;
\]%
\[
E\left(  \nu\left(  \overline{U}(j)\right)  |\overline{a}(j),\overline
{z}(j),\overline{x}\left(  j\right)  \right)  =0\Longrightarrow\nu\left(
\overline{U}(j)\right)  =0;
\]
and that there exist a function $H_{j}(\overline{a})=h_{j}\left(  \overline
{a},\overline{W}(j),a_{0}\right)  $ such that
\[
E\left(  H_{j+1}\left(  \overline{a}\right)  |\overline{z}\left(  j\right)
,\overline{a},\overline{x}\left(  j\right)  \right)  =E\left(  H_{j}\left(
\overline{a}\right)  |\overline{z}\left(  j\right)  ,\overline{a},\overline
{x}\left(  j\right)  \right)
\]
generalizations of conditions $\left(  \ref{long NC1}\right)  $-$\left(
\ref{confounding bridge long2}\right)  $; then we have that
\begin{align*}
E\left\{  Y_{\overline{a}}|\overline{x}(j),\overline{a}\left(  j-1\right)
\right\}   &  =E\left\{  h_{j}\left(  \overline{a},\overline{W}(j),\overline
{X}(j)\right)  |\overline{x}(j),\overline{a}\left(  j-1\right)  \right\}  ;\\
E\left\{  Y_{\overline{a}}|\overline{x}(j),\overline{u}(j),\overline{a}\left(
j-1\right)  \right\}   &  =E\left\{  h_{j}\left(  \overline{a},\overline
{W}(j),\overline{X}(j)\right)  |\overline{x}(j),\overline{u}\left(  j\right)
,\overline{a}\left(  j-1\right)  \right\}
\end{align*}
and
\[
E\left\{  Y_{\overline{a}}\right\}  =E\left\{  h_{0}\left(  \overline
{a},W(0),X(0)\right)  \right\}
\]

To prove the result, consider $j=J-1,$ then we have that
\begin{align*}
0 &  =E\left(  Y-h_{J-1}\left(  \overline{a},\overline{W},\overline{x}\right)
|\overline{a},\overline{z},\overline{x}\right)  \\
&  =E\left(  E\left\{  Y_{a(J-1)}-h_{J-1}\left(  \overline{a},\overline
{W},\overline{x}\right)  |\overline{U},\overline{a},\overline{z},\overline
{x}\right\}  |\overline{a},\overline{z},\overline{x}\right)  \\
&  \Rightarrow0=E\left\{  Y_{a(J-1)}-h_{J-1}\left(  \overline{a},\overline
{W},\overline{x}\right)  |\overline{U},\overline{a}\left(  J-2\right)
,\overline{x}\right\}  ;
\end{align*}
therefore
\[
E\left\{  Y_{a(J-1)}|\overline{a}\left(  J-2\right)  ,\overline{x}\right\}
=E\left\{  h_{J-1}\left(  \overline{a},\overline{W},\overline{x}\right)
|\overline{a}\left(  J-2\right)  ,\overline{x}\right\}
\]
Next,
\begin{align*}
0 &  =E\left(  h_{J-1}\left(  \overline{a},\overline{W},\overline{X}\right)
-h_{J-2}\left(  \overline{a},\overline{W}(J-2),\overline{X}(J-2)\right)
|\overline{a}\left(  J-2\right)  ,\overline{z}(J-2),\overline{x}(J-2)\right)
\\
&  \Rightarrow0=E\left[  \left.  E\left\{  \left.
\begin{array}
[c]{c}%
h_{J-1}\left(  \overline{a},\overline{W},\overline{X}\right)  \\
-h_{J-2}\left(  \overline{a},\overline{W}(J-2),\overline{X}(J-2)\right)
\end{array}
\right\vert \overline{U},\overline{z}(J-2),\overline{x}(J-2),\overline
{A}\right\}  \right\vert \overline{a}\left(  J-2\right)  ,\overline
{z}(J-2),\overline{x}(J-2)\right]  \\
&  \Rightarrow0=E\left[  E\left\{  h_{J-1}\left(  \overline{a},\overline
{W},\overline{X}\right)  -h_{J-2}\left(  \overline{a},\overline{W}%
(J-2),\overline{X}(J-2)\right)  |\overline{U},\overline{x},\overline
{A}\right\}  |\overline{a}\left(  J-2\right)  ,\overline{z}(J-2),\overline
{x}(J-2)\right]  \\
&  \Rightarrow0=E\left[  E\left\{  Y_{a(J-1)}-h_{J-2}\left(  \overline
{a},\overline{W}(J-2),\overline{X}(J-2)\right)  |\overline{U},\overline
{x},\overline{A}\right\}  |\overline{a}\left(  J-2\right)  ,\overline
{z}(J-2),\overline{x}(J-2)\right]  \\
&  \Rightarrow0=E\left[  \left.  E\left\{  \left.
\begin{array}
[c]{c}%
Y_{\overline{a}}\\
-h_{J-2}\left(  \overline{a},\overline{W}(J-2),\overline{X}(J-2)\right)
\end{array}
\right\vert \left.
\begin{array}
[c]{c}%
\overline{U},\overline{x},\\
\overline{a}\left(  J-2\right)  ,\overline{z}(J-2)
\end{array}
\right.  \right\}  \right\vert \left.
\begin{array}
[c]{c}%
\overline{a}\left(  J-2\right)  \\
,\overline{z}(J-2),\overline{x}(J-2)
\end{array}
\right.  \right]  \\
&  \Rightarrow0=E\left[  \left.  E\left\{  \left.
\begin{array}
[c]{c}%
Y_{\overline{a}}\\
-h_{J-2}\left(  \overline{a},\overline{W}(J-2),\overline{X}(J-2)\right)
\end{array}
\right\vert \left.
\begin{array}
[c]{c}%
\overline{U}(J-2),\overline{x}(J-2),\\
\overline{a}\left(  J-2\right)  ,\overline{z}(J-2)
\end{array}
\right.  \right\}  \right\vert \left.
\begin{array}
[c]{c}%
\overline{a}\left(  J-2\right)  ,\\
\overline{z}(J-2),\overline{x}(J-2)
\end{array}
\right.  \right]  \\
&  \Rightarrow0=E\left\{  Y_{\overline{a}}-h_{J-2}\left(  \overline
{a},\overline{W}(J-2),\overline{X}(J-2)\right)  |\overline{U}(J-2),\overline
{x}(J-2),\overline{a}\left(  J-3\right)  \right\}
\end{align*}
Therefore
\[
E\left\{  Y_{\overline{a}}|\overline{x}(J-2),\overline{a}\left(  J-3\right)
\right\}  =E\left\{  h_{J-2}\left(  \overline{a},\overline{W}(J-2),\overline
{X}(J-2)\right)  |\overline{x}(J-2),\overline{a}\left(  J-3\right)  \right\}
\]
Repeating this argument for $j=J-3,....0,$ we arrive at
\[
E\left\{  Y_{\overline{a}}|\overline{x}(j),\overline{a}\left(  j-1\right)
\right\}  =E\left\{  h_{j}\left(  \overline{a},\overline{W}(j),\overline
{X}(j)\right)  |\overline{x}(j),\overline{a}\left(  j-1\right)  \right\}
\]
and
\[
E\left\{  Y_{\overline{a}}-h_{0}\left(  \overline{a},W(0),X(0)\right)
\right\}  =0
\]
proving the result.

\bigskip

\noindent\underline{Closed form expression of $\mu\left(  A,\eta\right)  $ for
binary $Y$}: Suppose that $Y$ is binary and $W$ is a continuous scalar
variable. Further suppose that
\begin{align*}
W &  =\left(  1,Z^{\prime},A,X^{\prime}\right)  \Theta+\varepsilon_{W};\\
&  \varepsilon_{W}\amalg\left(  A,Z,X\right)  ;\\
\varepsilon_{W} &  \sim f_{\varepsilon_{W}}\text{ ;}%
\end{align*}
next, suppose that
\begin{align*}
g\left(  h\left(  w,A,X;\eta\right)  \right)   &  =\left(  1,W^{\prime
},A,X^{\prime}\right)  \eta\\
&  g\text{ is a known link function }%
\end{align*}
then we have that%
\begin{align*}
\mu\left(  A;\eta\right)   &  =%
{\displaystyle\sum\limits_{w}}
f_{W|Z,A}(w)h\left(  w,A,X;\eta\right)  \\
&  =%
{\displaystyle\sum\limits_{\varepsilon_{W}}}
f_{\varepsilon_{W}}\left(  \varepsilon_{W}\right)  g^{-1}\left(  \left(
1,\left(  1,Z^{\prime},A,X^{\prime}\right)  \Theta+\varepsilon_{W}%
,A,X^{\prime}\right)  \eta\right)  ;
\end{align*}
Suppose finally that one specifies $f_{\varepsilon_{W}}$ to match the bridge
distribution function of the link function $g^{20},$ then one can show that
\begin{align*}
\mu\left(  A;\eta,\Theta\right)   &  =g^{-1}\left(  \left(  1,\left(
1,Z^{\prime},A,X^{\prime}\right)  \Theta,A,X^{\prime}\right)  \eta^{\ast
}\right)  \\
\eta^{\ast} &  =\eta\times\phi
\end{align*}
with $0<\phi<1^{20}.$ The form of $\phi$ depends on the bridge distribution
function for the link $g.$ For instance, for $g$ the probit link, we have that
$f_{\varepsilon_{W}}$ is a zero mean Gaussian density with variance
$\sigma^{2}$ and $\phi=\left\{  1+\sigma^{2}\eta_{w}^{2}\right\}  ^{-1/2}$
and
\[
\mu\left(  A,\eta;\Theta\right)  =g^{-1}\left(  \left(  1,\left(  1,Z^{\prime
},A,X^{\prime}\right)  \Theta,A,X^{\prime}\right)  \eta\left\{  1+\sigma
^{2}\eta_{w}^{2}\right\}  ^{-1/2}\right)
\]
All parameters can be estimated by maximizing the log-pseudo-likelihood
function
\[
E_{n}\left\{  Y\log\left(  \mu\left(  A,\eta;\Theta\right)  \right)  +\left(
1-Y\right)  \log\left(  1-\mu\left(  A,\eta;\Theta\right)  \right)  \right\}
+\log f_{\varepsilon_{W}}\left(  \varepsilon_{W};\Theta,\sigma^{2}\right)
\]
In case $g$ is logit link, the above log likelihood is modified by setting
$\phi=\left\{  1+3\sigma^{2}\eta_{w}^{2}/\pi^{2}\right\}  ^{-1/2}$ and
$f_{\varepsilon_{W}}\left(  \varepsilon_{W};\Theta,\delta\right)  $ the
logistic bridge function $B_{l}\left(  0,\delta\right)  $ of Wang and
Louis$^{25}.$ In case $W$ is multivariate with both continuous and discrete
components we factorize $f_{W|Z,A}(W)=$ $f_{Wc|W_{d},Z,A}(W_{c})$
$f_{W_{d}|Z,A}(W_{d})$ where $W_{c}=\left(  W_{c,1},...,W_{c,d_{c}}\right)  $
are continuous components of $W$ and $W_{d}$ are discrete components. It is
then convenient to take $g$ as probit link function and $W_{c}|W_{d},Z,A,X$ as
multivariate Gaussian with mean $0$ and variance-covariance matrix $\Sigma$,
in which case%
\[
\mu\left(  A;\eta,\Theta\right)  =%
{\displaystyle\sum\limits_{w_{d}}}
f_{W_{d}|Z,A,X}(w_{d})g^{-1}\left(  \left(  1,\left(  1,Z^{\prime}%
,A,X^{\prime},W_{d}^{^{\prime}}\right)  \Theta,W_{d}^{^{\prime}},A,X^{\prime
}\right)  \eta\left\{  1+\eta_{w,c}^{\prime}\Sigma\eta_{w,c}\right\}
^{-1/2}\right)
\]

\bigskip

\noindent\underline{Generalization of Proximal recursive least squares
algorithm:}\noindent

\noindent Step 1: For user-specified functions $\mathrm{c}_{z,j}\left(
\overline{Z}\left(  j)\right)  \right)  ,\mathrm{c}_{a,j}\left(  \overline
{A}\right)  ,\mathrm{c}_{x,j}\left(  \overline{X}\left(  0\right)  \right)  $
fit the multivariate linear regression
\[
\mathrm{c}_{w,j}\left(  \overline{W}(j)\right)  =\left(  1,\mathrm{c}%
_{z,j}\left(  \overline{Z}\left(  j\right)  \right)  ^{\prime},\mathrm{c}%
_{a,j}\left(  \overline{A}\right)  ^{\prime},\mathrm{c}_{x,j}\left(
\overline{X}\left(  j\right)  \right)  ^{\prime},X^{\prime}(0)\right)
\Theta_{j}+\mathbf{\varepsilon}_{W,j}%
\]
$j=J-1,...,0$ by applying least-squares separately to each entry of vector
$\mathrm{c}_{w,j}\left(  \overline{W}(j)\right)  $, and let
\[
\widehat{\mathrm{c}}_{w,j}=\left(  1,\mathrm{c}_{z,j}\left(  \overline
{Z}\left(  j\right)  \right)  ^{\prime},\mathrm{c}_{a,j}\left(  \overline
{A}\right)  ^{\prime},\mathrm{c}_{x,j}\left(  \overline{X}\left(  j\right)
\right)  ^{\prime},X^{\prime}(0)\right)  \widehat{\Theta}_{j}%
\]
denote its fitted values;

\noindent Step 2:Let $\widehat{H}_{J}=Y$ and for $j=J-1,...,0,$ fit the linear
regression of
\[
\widehat{H}_{j+1}=\left(  1,\mathrm{c}_{a,j}\left(  \overline{A}\right)
^{\prime},\widehat{\mathrm{c}}_{w,j}^{\prime},\mathrm{c}_{x,j}\left(
\overline{X}\left(  j\right)  \right)  ^{\prime},X^{\prime}(0)\right)
\eta_{j}+\mathbf{\varepsilon}_{Y,j}%
\]
by least-squares where we note that $\widehat{\mathrm{c}}_{w,j}\left(
\overline{W}(j)\right)  $ has been substituted in for $\mathrm{c}_{w,j}\left(
\overline{W}(j)\right)  ,$ and let
\[
\widehat{H}_{j}\left(  \overline{A}\right)  =\left(  1,\mathrm{c}_{a,j}\left(
\overline{A}\right)  ^{\prime},\mathrm{c}_{w,j}\left(  \overline{W}(j)\right)
^{\prime},\mathrm{c}_{x,j}\left(  \overline{X}\left(  j\right)  \right)
^{\prime},X^{\prime}(0)\right)  \widehat{\eta}_{j}%
\]

\noindent Step 3:Evaluate%
\[
\widehat{\beta}\left(  \overline{a}\right)  =E_{n}\left\{  \widehat{H}%
_{0}\left(  \overline{a}\right)  \right\}  =\left(  1,\mathrm{c}_{a,0}\left(
\overline{a}\right)  ^{\prime},E_{n}\left\{  W^{\prime}\left(  0\right)
\right\}  ,E_{n}\left\{  X^{\prime}(0)\right\}  \right)  \widehat{\eta}_{0}.
\]

Next we show that $\widehat{\beta}\left(  \overline{a}\right)  $ is consistent
for $\beta\left(  \overline{a}\right)  $ provided that
\[
E\left\{  H_{j}\left(  \overline{a}\right)  |\overline{Z}\left(  j\right)
,\overline{a}\left(  j\right)  ,\overline{X}\left(  j\right)  \right\}
=E\left\{  \left(  1,\mathrm{c}_{a,j}\left(  \overline{a}\right)  ^{\prime
},\mathrm{c}_{w,j}\left(  \overline{W}(j)\right)  ^{\prime},\mathrm{c}%
_{x,j}\left(  \overline{X}\left(  j\right)  \right)  ^{\prime},X^{\prime
}(0)\right)  \eta_{j}|\overline{Z}\left(  j\right)  ,\overline{a}\left(
j\right)  ,\overline{X}\left(  j\right)  \right\}  ;
\]
even if
\[
\mathrm{c}_{w,j}\left(  \overline{W}(j)\right)  =\left(  1,\mathrm{c}%
_{z,j}\left(  \overline{Z}\left(  j\right)  \right)  ^{\prime},\mathrm{c}%
_{a,j}\left(  \overline{A}\right)  ^{\prime},\mathrm{c}_{x,j}\left(
\overline{X}\left(  j\right)  \right)  ^{\prime},X^{\prime}(0)\right)
\Theta_{j}+\mathbf{\varepsilon}_{W,j}%
\]
is misspecified. To prove this result, it suffices to note that
\begin{align}
0  &  =E_{n}\left\{  \left(
\begin{array}
[c]{c}%
1\\
\mathrm{c}_{a,J-1}\left(  \overline{A}\right) \\
\widehat{\mathrm{c}}_{w,J-1}\\
\mathrm{c}_{x,J-1}\left(  \overline{X}\left(  J-1\right)  \right) \\
X^{\prime}(0)
\end{array}
\right)  \left(  Y-\left(  1,\mathrm{c}_{a,J-1}\left(  \overline{A}\right)
^{\prime},\widehat{\mathrm{c}}_{w,J-1}^{\prime},\mathrm{c}_{x,J-1}\left(
\overline{X}\left(  J-1\right)  \right)  ^{\prime},X^{\prime}(0)\right)
\widehat{\eta}_{J-1}\right)  \right\} \\
&  =E_{n}\left\{  \left(
\begin{array}
[c]{c}%
1\\
\mathrm{c}_{a,J-1}\left(  \overline{A}\right) \\
\widehat{\mathrm{c}}_{w,J-1}\\
\mathrm{c}_{x,J-1}\left(  \overline{X}\left(  J-1\right)  \right) \\
X^{\prime}(0)
\end{array}
\right)  \left(  Y-\left(  1,\mathrm{c}_{a,J-1}\left(  \overline{A}\right)
^{\prime},\mathrm{c}_{w,J-1}\left(  \overline{W}(J-1)\right)  ^{\prime
},\mathrm{c}_{x,J-1}\left(  \overline{X}\left(  J-1\right)  \right)  ^{\prime
},X^{\prime}(0)\right)  \widehat{\eta}_{J-1}\right)  \right\}
\label{Moment equation 1}%
\end{align}
because
\[
0=E_{n}\left\{  \left(
\begin{array}
[c]{c}%
1\\
\mathrm{c}_{a,J-1}\left(  \overline{A}\right) \\
\widehat{\mathrm{c}}_{w,J-1}\\
\mathrm{c}_{x,J-1}\left(  \overline{X}\left(  J-1\right)  \right) \\
X^{\prime}(0)
\end{array}
\right)  \left(  \widehat{\mathrm{c}}_{w,J-1}-\mathrm{c}_{w,J-1}\left(
\overline{W}(J-1)\right)  \right)  \widehat{\eta}_{J-1}\right\}
\]
by virtue of $\widehat{\mathrm{c}}_{w,J-1}\left(  \overline{W}(J-1)\right)  $
being the least-square projection of $\mathrm{c}_{w,J-1}\left(  \overline
{W}(J-1)\right)  $ onto$\,$%
\[
\left(  1,\mathrm{c}_{a,J-1}\left(  \overline{A}\right)  ,\mathrm{c}%
_{z,j}\left(  \overline{Z}\left(  j\right)  \right)  ^{\prime},\mathrm{c}%
_{x,J-1}\left(  \overline{X}\left(  J-1\right)  \right)  ,X^{\prime
}(0)\right)
\]
which spans
\[
\left(  1,\mathrm{c}_{a,J-1}\left(  \overline{A}\right)  ,\widehat{\mathrm{c}%
}_{w,J-1}\left(  \overline{W}(J-1)\right)  ,\mathrm{c}_{x,J-1}\left(
\overline{X}\left(  J-1\right)  \right)  ,X^{\prime}(0)\right)  .
\]
Equation $\left(  \ref{Moment equation 1}\right)  $ yields a consistent
estimator of $\eta_{J-1}$ because $E(Y-H_{J-1}\left(  \eta_{J-1}\right)
|\overline{A},\overline{X},\overline{Z})=0.$ Likewise, for any $j<J-1$ we have
that
\begin{align*}
0  &  =E_{n}\left\{  \left(
\begin{array}
[c]{c}%
1\\
\mathrm{c}_{a,j}\left(  \overline{A}\right) \\
\widehat{\mathrm{c}}_{w,j}\\
\mathrm{c}_{x,j}\left(  \overline{X}\left(  j\right)  \right) \\
X^{\prime}(0)
\end{array}
\right)  \left(  \widehat{H}_{j+1}-\left(  1,\mathrm{c}_{a,j}\left(
\overline{A}\right)  ^{\prime},\widehat{\mathrm{c}}_{w,j}^{\prime}%
,\mathrm{c}_{x,j}\left(  \overline{X}\left(  j\right)  \right)  ^{\prime
},X^{\prime}(0)\right)  \widehat{\eta}_{j}\right)  \right\} \\
&  =E_{n}\left\{  \left(
\begin{array}
[c]{c}%
1\\
\mathrm{c}_{a,j}\left(  \overline{A}\right) \\
\widehat{\mathrm{c}}_{w,j}\\
\mathrm{c}_{x,j}\left(  \overline{X}\left(  j\right)  \right) \\
X^{\prime}(0)
\end{array}
\right)  \left(  \widehat{H}_{j+1}-\left(  1,\mathrm{c}_{a,j}\left(
\overline{A}\right)  ^{\prime},\mathrm{c}_{w,j}\left(  \overline{W}(j)\right)
^{\prime},\mathrm{c}_{x,j}\left(  \overline{X}\left(  j\right)  \right)
^{\prime},X^{\prime}(0)\right)  \widehat{\eta}_{j}\right)  \right\}
\end{align*}
because
\[
0=E_{n}\left\{  \left(
\begin{array}
[c]{c}%
1\\
\mathrm{c}_{a,j}\left(  \overline{A}\right) \\
\widehat{\mathrm{c}}_{w,j}\\
\mathrm{c}_{x,j}\left(  \overline{X}\left(  j\right)  \right) \\
X^{\prime}(0)
\end{array}
\right)  \left(  \mathrm{c}_{w,j}\left(  \overline{W}(j)\right)  ^{\prime
}-\widehat{\mathrm{c}}_{w,j}^{\prime}\right)  \widehat{\eta}_{j}\right\}
\]
by virtue of $\widehat{\mathrm{c}}_{w,j}$ being the least-square projection of
$\mathrm{c}_{w,j}\left(  \overline{W}(j)\right)  $ onto$\,\ $%
\[
\left(  1,\mathrm{c}_{a,j}\left(  \overline{A}\right)  ,\mathrm{c}%
_{z,j}\left(  \overline{Z}\left(  j\right)  \right)  ^{\prime},\mathrm{c}%
_{x,j}\left(  \overline{X}\left(  j\right)  \right)  ,X^{\prime}(0)\right)
\]
which spans
\[
\left(  1,\mathrm{c}_{a,j}\left(  \overline{A}\right)  ,\widehat{\mathrm{c}%
}_{w,j},\mathrm{c}_{x,j}\left(  \overline{X}\left(  j\right)  \right)
,X^{\prime}(0)\right)  .
\]

\clearpage
\renewcommand\thetable{A.\arabic{table}}\setcounter{table}{0}
\begin{landscape}
	\begin{table}[h]
		\caption{Examples of graphs for $Z,A,U$ relationships and for $W,Y,U$ relationships. The two pieces of graphs can be combined in to a  directed acyclic graph that encodes the assumptions on proxy variables of types b and c. Grey colored graphs are invalid due to violation of key assumptions.}\label{example_ZW}
		\centering{\small
			\renewcommand{\arraystretch}{1.45}
			\begin{tabular}{|c|c|c|c|}
				\hline 
				\multicolumn{4}{|c|}{Examples of graphs for $Z,A,U$ relationships}\tabularnewline
				\hline 
				& $Z\rightarrow A$ (pre-treatment) & $A\rightarrow Z$  (post-treatment) & $Z\ind A$ \tabularnewline
				\hline 
				No arrow between & Instrumental variable (IV) & Violate (\ref{completeness condition 1}) & Violate  (\ref{completeness condition 1})  \tabularnewline
				$U$ and $Z$  & \multirow{2}{*}{\resizebox{1.25in}{0.4in}{
						\begin{tikzpicture}
						\tikzset{line width=1pt,inner sep=5pt,
							ell/.style={draw, inner sep=1.5pt,line width=1pt}}
						\node[shape=ellipse,ell] (A) at (1,-1.5) {$A$};
						\node[shape=ellipse,ell] (U) at (2,-0.5) {$U,X$};
						\node[shape=ellipse,ell] (Y) at (3,-1.5) {$Y$};
						\node[shape=circle,ell,inner sep=2.2pt] (Z) at (-1.1,-1.5) {$Z$};
						\foreach \from/\to in {U/A,U/Y,A/Y,Z/A}
						\draw[-stealth,line width=0.5pt] (\from) -- (\to);
						\end{tikzpicture}} } & \multirow{2}{*}{\resizebox{1.25in}{0.4in}{
						\begin{tikzpicture}
						\tikzset{line width=1pt,inner sep=5pt, color=black!30,
							ell/.style={draw, inner sep=1.5pt,line width=1pt,color=black!30}}
						\node[shape=ellipse,ell] (A) at (1,-1.5) {$A$};
						\node[shape=ellipse,ell] (U) at (2,-0.5) {$U,X$};
						\node[shape=ellipse,ell] (Y) at (3,-1.5) {$Y$};
						\node[shape=circle,ell,inner sep=2.2pt] (Z) at (-1.1,-1.5) {$Z$};
						\foreach \from/\to in {U/A,U/Y,A/Y,A/Z}
						\draw[-stealth,line width=0.5pt,color=black!30] (\from) -- (\to);
						\end{tikzpicture}} } & \multirow{2}{*}{\resizebox{1.25in}{0.4in}{
						\begin{tikzpicture}
						\tikzset{line width=1pt,inner sep=5pt, color=black!30,
							ell/.style={draw, inner sep=1.5pt,line width=1pt,color=black!30}}
						\node[shape=ellipse,ell] (A) at (1,-1.5) {$A$};
						\node[shape=ellipse,ell] (U) at (2,-0.5) {$U,X$};
						\node[shape=ellipse,ell] (Y) at (3,-1.5) {$Y$};
						\node[shape=circle,ell,inner sep=2.2pt] (Z) at (-1.1,-1.5) {$Z$};
						\foreach \from/\to in {U/A,U/Y,A/Y}
						\draw[-stealth,line width=0.5pt,color=black!30] (\from) -- (\to);
						\end{tikzpicture}} } \tabularnewline
				(may violate (\ref{completeness condition 1}))& & &  \tabularnewline
				\cline{1-4} 
				& Invalid IV & Post-treatment proxy of $U$ & Surrogate of $U$ \tabularnewline
				$U\rightarrow Z$ & \multirow{2}{*}{\resizebox{1.25in}{0.4in}{
						\begin{tikzpicture}
						\tikzset{line width=1pt,inner sep=5pt,
							ell/.style={draw, inner sep=1.5pt,line width=1pt}}
						\node[shape=ellipse,ell] (A) at (1,-1.5) {$A$};
						\node[shape=ellipse,ell] (U) at (2,-0.5) {$U,X$};
						\node[shape=ellipse,ell] (Y) at (3,-1.5) {$Y$};
						\node[shape=circle,ell,inner sep=2.2pt] (Z) at (-1.1,-1.5) {$Z$};
						\foreach \from/\to in {U/A,U/Y,A/Y,U/Z,Z/A}
						\draw[-stealth,line width=0.5pt] (\from) -- (\to);
						\end{tikzpicture}}} & \multirow{2}{*}{\resizebox{1.25in}{0.4in}{
						\begin{tikzpicture}
						\tikzset{line width=1pt,inner sep=5pt,
							ell/.style={draw, inner sep=1.5pt,line width=1pt}}
						\node[shape=ellipse,ell] (A) at (1,-1.5) {$A$};
						\node[shape=ellipse,ell] (U) at (2,-0.5) {$U,X$};
						\node[shape=ellipse,ell] (Y) at (3,-1.5) {$Y$};
						\node[shape=circle,ell,inner sep=2.2pt] (Z) at (-1.1,-1.5) {$Z$};
						\foreach \from/\to in {U/A,U/Y,A/Y,U/Z,A/Z}
						\draw[-stealth,line width=0.5pt] (\from) -- (\to);
						\end{tikzpicture}}} & \multirow{2}{*}{\resizebox{1.25in}{0.4in}{
						\begin{tikzpicture}
						\tikzset{line width=1pt,inner sep=5pt,
							ell/.style={draw, inner sep=1.5pt,line width=1pt}}
						\node[shape=ellipse,ell] (A) at (1,-1.5) {$A$};
						\node[shape=ellipse,ell] (U) at (2,-0.5) {$U,X$};
						\node[shape=ellipse,ell] (Y) at (3,-1.5) {$Y$};
						\node[shape=circle,ell,inner sep=2.2pt] (Z) at (-1.1,-1.5) {$Z$};
						\foreach \from/\to in {U/A,U/Y,A/Y,U/Z}
						\draw[-stealth,line width=0.5pt] (\from) -- (\to);
						\end{tikzpicture}}} \tabularnewline
				& & &  \tabularnewline
				\hline 
				& \multicolumn{3}{c|}{May violate (\ref{NC.2}) if there is $W\rightarrow U$} \tabularnewline
				$Z\rightarrow U$ & \multirow{2}{*}{\resizebox{1.25in}{0.4in}{
						\begin{tikzpicture}
						\tikzset{line width=1pt,inner sep=5pt,
							ell/.style={draw, inner sep=1.5pt,line width=1pt}}
						\node[shape=ellipse,ell] (A) at (1,-1.5) {$A$};
						\node[shape=ellipse,ell] (U) at (2,-0.5) {$U,X$};
						\node[shape=ellipse,ell] (Y) at (3,-1.5) {$Y$};
						\node[shape=circle,ell,inner sep=2.2pt] (Z) at (-1.1,-1.5) {$Z$};
						\foreach \from/\to in {U/A,U/Y,A/Y,Z/U,Z/A}
						\draw[-stealth,line width=0.5pt] (\from) -- (\to);
						\end{tikzpicture}}} & \multirow{2}{*}{\resizebox{1.25in}{0.4in}{
						\begin{tikzpicture}
						\tikzset{line width=1pt,inner sep=5pt,
							ell/.style={draw, inner sep=1.5pt,line width=1pt}}
						\node[shape=ellipse,ell] (A) at (1,-1.5) {$A$};
						\node[shape=ellipse,ell] (U) at (2,-0.5) {$U,X$};
						\node[shape=ellipse,ell] (Y) at (3,-1.5) {$Y$};
						\node[shape=circle,ell,inner sep=2.2pt] (Z) at (-1.1,-1.5) {$Z$};
						\foreach \from/\to in {U/A,U/Y,A/Y,Z/U,A/Z}
						\draw[-stealth,line width=0.5pt] (\from) -- (\to);
						\end{tikzpicture}}} & \multirow{2}{*}{\resizebox{1.25in}{0.4in}{
						\begin{tikzpicture}
						\tikzset{line width=1pt,inner sep=5pt,
							ell/.style={draw, inner sep=1.5pt,line width=1pt}}
						\node[shape=ellipse,ell] (A) at (1,-1.5) {$A$};
						\node[shape=ellipse,ell] (U) at (2,-0.5) {$U,X$};
						\node[shape=ellipse,ell] (Y) at (3,-1.5) {$Y$};
						\node[shape=circle,ell,inner sep=2.2pt] (Z) at (-1.1,-1.5) {$Z$};
						\foreach \from/\to in {U/A,U/Y,A/Y,Z/U}
						\draw[-stealth,line width=0.5pt] (\from) -- (\to);
						\end{tikzpicture}}} \tabularnewline
				& & & \tabularnewline
				\hline 
				\multicolumn{4}{|c|}{Examples of graphs for $W,Y,U$ relationships}\tabularnewline
				\hline 
				& $W\rightarrow Y(a)$ & $Y(a)\rightarrow W$ & $Y(a)\ind W\mid(U,X)$\tabularnewline
				& & Violate (\ref{NC.2}) & \tabularnewline
				\hline 
				No arrow between& Violate (\ref{completeness condition 2}) & Violate (\ref{NC.2}) and (\ref{completeness condition 2}) & Violate (\ref{completeness condition 2}) \tabularnewline
				$U$ and $W$ (violate & \multirow{2}{*}{\resizebox{1.25in}{0.4in}{
						\begin{tikzpicture}
						\tikzset{line width=1pt,inner sep=5pt, color=black!30,
							ell/.style={draw, inner sep=1.5pt,line width=1pt,color=black!30}}
						\node[shape=ellipse,ell] (A) at (1,-1.5) {$A$};
						\node[shape=ellipse,ell] (U) at (2,-0.5) {$U,X$};
						\node[shape=ellipse,ell] (Y) at (3,-1.5) {$Y$};
						\node[shape=circle,ell] (W) at (5,-1.5) {$W$};
						\foreach \from/\to in {U/A,U/Y,A/Y,W/Y}
						\draw[-stealth,line width=0.5pt,color=black!30] (\from) -- (\to);
						\end{tikzpicture}}} & \multirow{2}{*}{\resizebox{1.25in}{0.4in}{
						\begin{tikzpicture}
						\tikzset{line width=1pt,inner sep=5pt, color=black!30,
							ell/.style={draw, inner sep=1.5pt,line width=1pt,color=black!30}}
						\node[shape=ellipse,ell] (A) at (1,-1.5) {$A$};
						\node[shape=ellipse,ell] (U) at (2,-0.5) {$U,X$};
						\node[shape=ellipse,ell] (Y) at (3,-1.5) {$Y$};
						\node[shape=circle,ell] (W) at (5,-1.5) {$W$};
						\foreach \from/\to in {U/A,U/Y,A/Y,Y/W}
						\draw[-stealth,line width=0.5pt,color=black!30] (\from) -- (\to);
						\end{tikzpicture}}} & \multirow{2}{*}{\resizebox{1.25in}{0.4in}{
						\begin{tikzpicture}
						\tikzset{line width=1pt,inner sep=5pt, color=black!30,
							ell/.style={draw, inner sep=1.5pt,line width=1pt,color=black!30}}
						\node[shape=ellipse,ell] (A) at (1,-1.5) {$A$};
						\node[shape=ellipse,ell] (U) at (2,-0.5) {$U,X$};
						\node[shape=ellipse,ell] (Y) at (3,-1.5) {$Y$};
						\node[shape=circle,ell] (W) at (5,-1.5) {$W$};
						\foreach \from/\to in {U/A,U/Y,A/Y}
						\draw[-stealth,line width=0.5pt,color=black!30] (\from) -- (\to);
						\end{tikzpicture}}}\tabularnewline
				(\ref{completeness condition 2}) )& & & \tabularnewline
				\hline 
				& & Violate (\ref{NC.2}) & \tabularnewline
				$U\rightarrow W$ & \multirow{2}{*}{\resizebox{1.25in}{0.4in}{
						\begin{tikzpicture}
						\tikzset{line width=1pt,inner sep=5pt,
							ell/.style={draw, inner sep=1.5pt,line width=1pt}}
						\node[shape=ellipse,ell] (A) at (1,-1.5) {$A$};
						\node[shape=ellipse,ell] (U) at (2,-0.5) {$U,X$};
						\node[shape=ellipse,ell] (Y) at (3,-1.5) {$Y$};
						\node[shape=circle,ell] (W) at (5,-1.5) {$W$};
						\foreach \from/\to in {U/A,U/Y,A/Y,W/Y,U/W}
						\draw[-stealth,line width=0.5pt] (\from) -- (\to);
						\end{tikzpicture}}} & \multirow{2}{*}{\resizebox{1.25in}{0.4in}{
						\begin{tikzpicture}
						\tikzset{line width=1pt,inner sep=5pt, color=black!30,
							ell/.style={draw, inner sep=1.5pt,line width=1pt,color=black!30}}
						\node[shape=ellipse,ell] (A) at (1,-1.5) {$A$};
						\node[shape=ellipse,ell] (U) at (2,-0.5) {$U,X$};
						\node[shape=ellipse,ell] (Y) at (3,-1.5) {$Y$};
						\node[shape=circle,ell] (W) at (5,-1.5) {$W$};
						\foreach \from/\to in {U/A,U/Y,A/Y,Y/W,U/W}
						\draw[-stealth,line width=0.5pt,color=black!30] (\from) -- (\to);
						\end{tikzpicture}}} & \multirow{2}{*}{\resizebox{1.25in}{0.4in}{
						\begin{tikzpicture}
						\tikzset{line width=1pt,inner sep=5pt,
							ell/.style={draw, inner sep=1.5pt,line width=1pt}}
						\node[shape=ellipse,ell] (A) at (1,-1.5) {$A$};
						\node[shape=ellipse,ell] (U) at (2,-0.5) {$U,X$};
						\node[shape=ellipse,ell] (Y) at (3,-1.5) {$Y$};
						\node[shape=circle,ell] (W) at (5,-1.5) {$W$};
						\foreach \from/\to in {U/A,U/Y,A/Y,U/W}
						\draw[-stealth,line width=0.5pt] (\from) -- (\to);
						\end{tikzpicture}}}\tabularnewline
				& & & \tabularnewline
				\hline 
				& \multicolumn{3}{c|}{May violate (\ref{NC.2}) if there is $Z\rightarrow U$}\tabularnewline
				& & Violate (\ref{NC.2}) & \tabularnewline
				$W\rightarrow U$ & \multirow{2}{*}{\resizebox{1.25in}{0.4in}{
						\begin{tikzpicture}
						\tikzset{line width=1pt,inner sep=5pt,
							ell/.style={draw, inner sep=1.5pt,line width=1pt}}
						\node[shape=ellipse,ell] (A) at (1,-1.5) {$A$};
						\node[shape=ellipse,ell] (U) at (2,-0.5) {$U,X$};
						\node[shape=ellipse,ell] (Y) at (3,-1.5) {$Y$};
						\node[shape=circle,ell] (W) at (5,-1.5) {$W$};
						\foreach \from/\to in {U/A,U/Y,A/Y,W/Y,W/U}
						\draw[-stealth,line width=0.5pt] (\from) -- (\to);
						\end{tikzpicture}}} & \multirow{2}{*}{\resizebox{1.25in}{0.4in}{
						\begin{tikzpicture}
						\tikzset{line width=1pt,inner sep=5pt, color=black!30,
							ell/.style={draw, inner sep=1.5pt,line width=1pt,color=black!30}}
						\node[shape=ellipse,ell] (A) at (1,-1.5) {$A$};
						\node[shape=ellipse,ell] (U) at (2,-0.5) {$U,X$};
						\node[shape=ellipse,ell] (Y) at (3,-1.5) {$Y$};
						\node[shape=circle,ell] (W) at (5,-1.5) {$W$};
						\foreach \from/\to in {U/A,U/Y,A/Y,Y/W,W/U}
						\draw[-stealth,line width=0.5pt,color=black!30] (\from) -- (\to);
						\end{tikzpicture}}} & \multirow{2}{*}{\resizebox{1.25in}{0.4in}{
						\begin{tikzpicture}
						\tikzset{line width=1pt,inner sep=5pt,
							ell/.style={draw, inner sep=1.5pt,line width=1pt}}
						\node[shape=ellipse,ell] (A) at (1,-1.5) {$A$};
						\node[shape=ellipse,ell] (U) at (2,-0.5) {$U,X$};
						\node[shape=ellipse,ell] (Y) at (3,-1.5) {$Y$};
						\node[shape=circle,ell] (W) at (5,-1.5) {$W$};
						\foreach \from/\to in {U/A,U/Y,A/Y,W/U}
						\draw[-stealth,line width=0.5pt] (\from) -- (\to);
						\end{tikzpicture}}}\tabularnewline
				& & & \tabularnewline
				\hline 
			\end{tabular}
		}
	\end{table}
\end{landscape}

\clearpage
\begin{table}[ptb]
	\caption{Results from right heart catherization empirical application.}%
	\label{tbl:rhc}%
	\centering
	\begin{tabular}{ccccccccc}
		\hline 
		\multicolumn{2}{c}{} & \multicolumn{3}{c}{Recursive Proximal 2SLS} &  & \multicolumn{3}{c}{Ordinary Least Squares}\tabularnewline
		\cline{3-5} \cline{7-9} 
		\multicolumn{2}{c}{Variable} & Estimate & Std Err & p-value &  & Estimate & Std Err & p-value\tabularnewline
		\hline 
		& RHC & -1.80 & 0.43 & $<$0.001 &  & -1.25 & 0.28 & $<$0.001\tabularnewline
		\multirow{3}{*}{} & Age & 0.05 & 0.04 & 0.25 &  & -0.01 & 0.01 & 0.27\tabularnewline
		& Sex (female) & -1.10 & 1.13 & 0.33 &  & 0.49 & 0.25 & 0.05\tabularnewline
		& Race (black) & -0.89 & 1.05 & 0.40 &  & 0.39 & 0.34 & 0.25\tabularnewline
		\multirow{2}{*}{W }& Serum pH & -16.92 & 8.80 & 0.05 &  & 3.11 & 1.41 & 0.03\tabularnewline
		& Hematocrit & -1.01 & 0.69 & 0.14 &  & -0.03 & 0.02 & 0.11\tabularnewline
		\multirow{2}{*}{Z} & PaO2/(.01{*}FiO2) &  &  &  &  & 0.00 & 0.00 & 0.03\tabularnewline
		& PaCO2 &  &  &  &  & 0.04 & 0.01 & 0.00\tabularnewline
		\hline 
	\end{tabular}
\end{table}

\begin{table}[!htbp] \centering 
	\caption{Results from Methotrexate empirical application.} 
	\begin{tabular}{@{\extracolsep{5pt}}lc} 
		\\[-1.8ex]\hline 
		\hline \\[-1.8ex] 
		& \multicolumn{1}{c}{\textit{Dependent variable:}} \\ 
		\cline{2-2} 
		\\[-1.8ex] & jc \\ 
		\hline \\[-1.8ex] 
		mtxspan & $-$0.154$^{***}$ (0.057) \\ 
		& p = 0.008 \\ 
		fitted\_jc & 0.524$^{***}$ (0.097) \\ 
		& p = 0.00000 \\ 
		\hline \\[-1.8ex] 
		\textit{Note:}  & \multicolumn{1}{r}{$^{*}$p$<$0.1; $^{**}$p$<$0.05; $^{***}$p$<$0.01} \\ 
	\end{tabular}\\\vspace{0.3in}
\begin{tabular}{@{\extracolsep{5pt}}lc} 
	\\[-1.8ex]\hline 
	\hline \\[-1.8ex] 
	& \multicolumn{1}{c}{\textit{Dependent variable:}} \\ 
	\cline{2-2} 
	\\[-1.8ex] & h\_1 \\ 
	\hline \\[-1.8ex] 
	cum\_treatment & $-$0.188$^{***}$ (0.057) \\ 
	& p = 0.00000 \\ 
	\hline \\[-1.8ex] 
	\textit{Note:}  & \multicolumn{1}{r}{$^{*}$p$<$0.1; $^{**}$p$<$0.05; $^{***}$p$<$0.01}\\ 
\end{tabular} 
\end{table}


\begin{thebibliography}{99}                                                                                               %


\bibitem {1}Greenland, S. and Robins, J.M., 1986. Identifiability,
exchangeability, and epidemiological confounding. International Journal of
Epidemiology, 15(3), pp.413-419.

\bibitem {2}Hern\'{a}n MA, Robins JM (2020). Causal Inference: What If. Boca
Raton: Chapman \& Hall/CRC

\bibitem {3}Hern\'{a}n, M.A., Brumback, B.A. and Robins, J.M., 2002.
Estimating the causal effect of zidovudine on CD4 count with a marginal
structural model for repeated measures. Statistics in Medicine, 21(12), pp.1689-1709.

\bibitem {4}Woods, S.P., Moore, D.J., Weber, E. and Grant, I., 2009. Cognitive
neuropsychology of HIV-associated neurocognitive disorders. Neuropsychology
Review, 19(2), pp.152-168.

\bibitem {5}Heaton, R.K., Franklin, D.R., Ellis, R.J., McCutchan, J.A.,
Letendre, S.L., LeBlanc, S., Corkran, S.H., Duarte, N.A., Clifford, D.B.,
Woods, S.P. and Collier, A.C., 2011. HIV-associated neurocognitive disorders
before and during the era of combination antiretroviral therapy: differences
in rates, nature, and predictors. Journal of Neurovirology, 17(1), pp.3-16.

\bibitem {6}Folstein, MF; Folstein, SE; McHugh, PR (1975). ""Mini-mental
status". A practical method for grading the cognitive state of patients for
the clinician". Journal of Psychiatric Research. 12 (3): 189--98.
doi:10.1016/0022-3956(75)90026-6. PMID 1202204.

\bibitem {7}Tombaugh, Tom N.; McIntyre, Nancy J. (1992). "The Mini Mental
Status Examination: A comprehensive review". JAGS. 40 (9): 922--935.
doi:10.1111/j.1532-5415.1992.tb01992.x. PMID 1512391.

\bibitem {8}Toglia, J., Fitzgerald, K.A., O'Dell, M.W., Mastrogiovanni, A.R.
and Lin, C.D., 2011. The Mini-Mental State Examination and Montreal Cognitive
Assessment in persons with mild subacute stroke: relationship to functional
outcome. Archives of physical medicine and rehabilitation, 92(5), pp.792-798.

\bibitem {9}Rosenbaum, P.R. and Rubin, D.B., 1983. The central role of the
propensity score in observational studies for causal effects. Biometrika,
70(1), pp.41-55.

\bibitem {10}Hern\'{a}n, M.A. and Robins, J.M., 2006. Instruments for causal
inference: an epidemiologist's dream?. Epidemiology, pp.360-372.

\bibitem {11}Wang, L. and Tchetgen Tchetgen, E.J., 2018. Bounded, efficient
and multiply robust estimation of average treatment effects using instrumental
variables. Journal of the Royal Statistical Society. Series B, Statistical
methodology, 80(3), p.531.

\bibitem {12}Miao, W., Shi, X. and Tchetgen Tchetgen, E.J., 2018. A Confounding Bridge
Approach for Double Negative Control Inference on Causal Effects. arXiv
preprint arXiv:1808.04945.

\bibitem {13}Miao, W. and Tchetgen Tchetgen, E.J, 2017. Invited commentary:
bias attenuation and identification of causal effects with multiple negative
controls. American journal of epidemiology, 185(10), pp.950-953.

\bibitem {14}Sofer, T., Richardson, D.B., Colicino, E., Schwartz, J. and
Tchetgen Tchetgen, E.J., 2016. On negative outcome control of unobserved
confounding as a generalization of difference-in-differences. Statistical
science: a review journal of the Institute of Mathematical Statistics, 31(3), p.348.

\bibitem {15}Miao, W., Geng, Z. and Tchetgen Tchetgen, E.J., 2018. Identifying
causal effects with proxy variables of an unmeasured confounder. Biometrika,
105(4), pp.987-993.

\bibitem {16}Shi, X., Miao, W., Nelson, J.C. and Tchetgen Tchetgen, E.J.,
2020. Multiply robust causal inference with double-negative control adjustment
for categorical unmeasured confounding. Journal of the Royal Statistical
Society: Series B (Statistical Methodology).

\bibitem {17}Robins, James M. "Causal inference from complex longitudinal
data." Latent variable modeling and applications to causality. Springer, New
York, NY, 1997. 69-117.

\bibitem {18}Wooldridge, J.M., 2010. Econometric analysis of cross section and
panel data. MIT press.

\bibitem {19}Efron, B. and Tibshirani, R.J., 1994. An introduction to the
bootstrap. CRC press.

\bibitem {20}Connors, A.F., Speroff, T., Dawson, N.V., Thomas, C., Harrell,
F.E., Wagner, D., Desbiens, N., Goldman, L., Wu, A.W., Califf, R.M. and
Fulkerson, W.J., 1996. The effectiveness of right heart catheterization in the
initial care of critically III patients. Jama, 276(11), pp.889-897.

\bibitem {21}Choi, H.K., Hern\'{a}n, M.A., Seeger, J.D., Robins, J.M. and
Wolfe, F., 2002. Methotrexate and mortality in patients with rheumatoid
arthritis: a prospective study. The Lancet, 359(9313), pp.1173-1177.

\bibitem {22}Lipsitch, M., Tchetgen, E.T. and Cohen, T., 2010. Negative
controls: a tool for detecting confounding and bias in observational studies.
Epidemiology (Cambridge, Mass.), 21(3), p.383.

\bibitem {23}Shi, X., Miao, W. and Tchetgen Tchetgen, EJ. A Selective Review
of Negative Control Methods in Epidemiology. arXiv:2009.0564

\bibitem {24}Scharfstein, D.O., Rotnitzky, A. and Robins, J.M., 1999.
Adjusting for nonignorable drop-out using semiparametric nonresponse models.
with discussions and rejoinder. Journal of the American Statistical
Association, 94(448), pp.1096-1120.

\bibitem {25}Wang, Z. and Louis, T.A., 2003. Matching conditional and marginal
shapes in binary random intercept models using a bridge distribution function.
Biometrika, 90(4), pp.765-775.

\pagebreak
\end{thebibliography}
\end{document}